\documentclass[iop,tighten,appendixfloats,numberedappendix,apj]{emulateapj} %
\usepackage{ifpdf} 
\pdfoutput=1  %
\usepackage{verse}
\usepackage{graphicx}
\usepackage[usenames, dvipsnames]{xcolor}
\definecolor{lightgreen}{HTML}{228B22}
\definecolor{darkgreen}{HTML}{006400}
\usepackage{cmap} 
\usepackage[colorlinks, bookmarks, breaklinks, pdftitle={Unification of the Fundamental Plane and BH masses},pdfauthor={Remco C. E. van den Bosch}]{hyperref}
\hypersetup{linkcolor=black,citecolor=black,filecolor=black,urlcolor=darkgreen,pdfborderstyle={/S/U}}   
\usepackage[english]{babel}
\usepackage[activate=true,kerning=true,tracking=true,spacing=true,babel,final,protrusion=true]{microtype}  
\usepackage[yyyymmdd,hhmmss]{datetime}
\usepackage{nicefrac}
\usepackage{mathptmx}
\usepackage[T1]{fontenc}
\usepackage{amsmath}

\def\kms{km\,s$^{-1}$}
\def\Mbh{$M_\bullet$}
\def\Msun{M$_\odot$}
\def\Lsun{L$_\odot$}

\def\MLsunK{M$_\odot$/L${_{\odot,K}}$}
\newcommand{\Msigma}   {\ensuremath{M_\bullet}{--}\ensuremath{\sigma}}

\def\Mbulge{\mbox{\Mbh--M$_{{\mathrm{bul}}}$}}

\newcommand{\todo}[1]{} 

\usepackage{relsize}
\usepackage{longtable}
\newcommand{\appropto}{\mathrel{\vcenter{
  \offinterlineskip\halign{\hfil$##$\cr
    \propto\cr\noalign{\kern2pt}\sim\cr\noalign{\kern-2pt}}}}}

\def\BHMRname{BH--size--mass relation}
\def\BHLRname{BH--size--luminosity relation}

\clearpage{}%
\def\ngal{230}

\def\MLfunc{$M_\star/L_K= 0.10 \sigma_e^{0.45}$}
\def\bulgeRL{\log (R_e/ \mathrm{kpc}) =-7.09+0.68 (L_k/\mathrm{L}_\sun)}
\def\bulgeMM{\log (M_\bullet/\mathrm{M}_\sun) =-4.98+1.21\log (M_\star/\mathrm{M}_\sun)}
\def\diskRL{\log (R_e/ \mathrm{kpc}) =-3.39+0.36 (L_k/\mathrm{L}_\sun)}
\def\diskMM{\log (M_\bullet/\mathrm{M}_\sun) =-14.36+2.02\log (M_\star/\mathrm{M}_\sun)}
\def\eqmsigmatab{$ -4.00  \pm  0.51 $ & $  5.35  \pm  0.23$ & $\log \sigma_e$ &  & -- & ($  0.49  \pm  0.03$) \\}
\def\eqmsigma{\log \left (\frac{M_\bullet}{ \mathrm{M_\sun}} \right)= (  8.32  \pm  0.04) + (   5.35  \pm  0.23) \log \left (\frac{\sigma_e}{ \mathrm{200 km\,s}^{-1}} \right) }
\def\eqmsigmaeps{($\epsilon=  0.49  \pm  0.03$)}
\def\msigmaapprox{$M_\bullet \propto \sigma_e^{5.4}$}
\def\msigmaapproxerr{$M_\bullet \propto \sigma_e^{5.4\pm0.2}$}
\def\eqbhlrtab{$-30.47  \pm  1.98$ & $  3.66  \pm  0.19$ & $\log L_k$ & $ -3.42  \pm  0.26$ & $\log R_e$ & ($  0.57  \pm  0.04$) \\}
\def\eqbhlr{\log \left (\frac{M_\bullet}{ \mathrm{M_\sun}} \right)= (  7.37  \pm  0.06)  \\ + (   3.66  \pm  0.19) \log \left (\frac{L_\star}{ 10^{11} \mathrm{L_\sun}} \right) \\ + ( -3.42  \pm  0.26) \log \left (\frac{R_e}{ 5 \mathrm{kpc}} \right)}
\def\eqbhlreps{($\epsilon=  0.57  \pm  0.04$)}
\def\eqbhmrtab{$-22.58  \pm  1.42$ & $  2.91  \pm  0.14$ & $\log M_\star$ & $ -2.77  \pm  0.22$ & $\log R_e$ & ($  0.52  \pm  0.03$) \\}
\def\eqbhmr{\log \left (\frac{M_\bullet}{ \mathrm{M_\sun}} \right)= (  7.48  \pm  0.05)  \\ + (   2.91  \pm  0.14) \log \left (\frac{M_\star}{ 10^{11} \mathrm{M_\sun}} \right) \\ + ( -2.77  \pm  0.22) \log \left (\frac{R_e}{ 5 \mathrm{kpc}} \right)}

\def\bhMRapprox{$M_\bullet \propto ( \nicefrac{M_\star}{R_e})^{2.9} $}

\def\eqfp{\log \left (\frac{\sigma_e}{ \mathrm{km\,s}^{-1}} \right)= (  2.11  \pm  0.01)  \\ + (   0.71  \pm  0.03) \log \left (\frac{L_\star}{ 10^{11} \mathrm{L_\sun}} \right) \\ + ( -0.72  \pm  0.05) \log \left (\frac{R_e}{ 5 \mathrm{kpc}} \right)}
\def\eqfpeps{($\epsilon=  0.07  \pm  0.01$)}
\def\fpLk{$L_k \propto R_e^{  1.01} \sigma_e^{  1.40}$}
\def\fptilt{$ R_e^{ -0.01} \sigma_e^{  0.60}$}

\def\VTestimator{G M_\star=  9.5 \sigma_e^{ 1.85} R_e^{ 1.01}}
\def\VTkappa{  9.5}
\def\eqMbhMtab{$ -9.84  \pm  1.11 $ & $  1.60  \pm  0.10$ & $\log M_\star$ &  & -- & ($  0.76  \pm  0.05$) \\}

\def\eqMbhLktab{$-11.06  \pm  1.38 $ & $  1.70  \pm  0.12$ & $\log L_k$ &  & -- & ($  0.84  \pm  0.05$) \\}

\def\eqMbhLkeps{($\epsilon=  0.84  \pm  0.05$)}
\def\eqmfixlrtab{$-31.55  \pm  1.92 $ & $  3.78  \pm  0.18$ & $\log \nicefrac{L_k}{R_e}$ &  & -- & ($  0.57  \pm  0.04$) \\}

\def\bhmasssize{$M_\bullet \propto (\nicefrac{L_k}{R_e})^{3.8}$}
\def\fancyeq{\log(M_\bullet) \propto \left( \frac{L_k^{  3.66}}{R_e^{  3.42}} \right) \appropto\left( \sigma_e^{  1.40}  \right)^{  3.78} \propto \sigma_e^{  5.29}}
\def\eqmsigmaRetab{$ -3.99  \pm  0.52$ & $  5.35  \pm  0.24$ & $\log \sigma_e$ & $ -0.01  \pm  0.13$ & $\log R_e$ & ($  0.49  \pm  0.03$) \\}

\def\eqmsigmaLktab{$ -4.94  \pm  0.96$ & $  5.07  \pm  0.32$ & $\log \sigma_e$ & $  0.14  \pm  0.12$ & $\log L_k$ & ($  0.49  \pm  0.03$) \\}

\def\eqmsigmaCtab{$ -3.96  \pm  0.51$ & $  5.10  \pm  0.28$ & $\log \sigma_e$ & $  0.10  \pm  0.07$ & $C_{28}$ & ($  0.49  \pm  0.03$) \\}

\clearpage{}%

\shortauthors{Remco C. E. van den Bosch} 
\shorttitle{The \BHLRname}
\submitted{Accepted by ApJ. In original form on February 29th 2016. Third revision submitted on June 20rd. Current revision \today.}  

\begin{document}

\title{Unification of the Fundamental Plane and Super-Massive Black Holes Masses}
\author{Remco C. E. van den Bosch \href{mailto:bosch@mpia.de}{(bosch@mpia.de)}%
}

 \begin{abstract}
According to the Virial Theorem, all gravitational systems in equilibrium sit on a plane in the 3D parameter space defined by their mass, size and second moment of the velocity tensor. While these quantities cannot be directly observed, there are suitable proxies: the luminosity $L_k$, half-light radius $R_e$ and dispersion $\sigma_e$. These proxies indeed lie on a very tight Fundamental Plane (FP).  How do the black holes in the centers of galaxies relate to the FP? Their masses are known to exhibit no strong correlation with total galaxy mass, but they do correlate weakly with bulge mass (when present), and extremely well with the velocity dispersion through the \msigmaapprox\ relation. These facts together imply that a tight plane must also exist defined by black hole mass, total galaxy mass and size. Here I show that this is indeed the case using a heterogeneous set of \ngal\  black holes. The sample  includes BHs from zero to 10 billion solar masses and host galaxies ranging from low surface brightness dwarfs, through bulge-less disks, to brightest cluster galaxies. The resulting \BHLRname\ \bhmasssize\ has the same amount of scatter as the \Msigma\ relation and is aligned with the galaxy FP, such that it is just a re-projection of $\sigma_e$.  The inferred \BHMRname\ is \bhMRapprox. These relationships are universal and extend to galaxies without bulges. This  implies that the black hole is primarily correlated with its global velocity dispersion and not with the properties of the bulge.  I show that the classical bulge--mass relation is a projection of the \Msigma\ relation.  When the velocity dispersion  cannot be  measured (at high-$z$ or low dispersions),  the  \BHMRname\  should be used as a proxy for black hole mass in favor of just galaxy or bulge mass. \\ Table and code available at \url{https://github.com/remcovandenbosch/Black-Hole-Mass-compilation}.
\end{abstract} 

\section{Dynamical scaling relations}

\noindent The Virial Theorem states that all gravitational systems in equilibrium sit on a plane $GM_{\nicefrac{1}{2}}= R_{\nicefrac{1}{2}}\sigma_{\nicefrac{1}{2}}^{2}$ in the 3D parameter space defined by half of the mass, the size and the second moment of the velocity tensor of those systems. While we cannot directly observe these quantities, we do have suitable proxies: the luminosity, half-light radius $R_e$ and stellar velocity dispersion $\sigma_e$. Together these three variables define the Fundamental Plane \citep{djorgovski87,dressler87} of early-type galaxies as is expected from the Virial Theorem. There are several empirical relationships that are understood to be projections of the Fundamental Plane: ellipticals follow the Faber-Jackson \citeyearpar{faber76} and Kormendy \citeyearpar{kormendy77a} relations, and all early-type galaxies lie on a narrow red sequence \citep{chen10c}. Spiral galaxies follow different dynamical scaling relations, including the Tully-Fisher \citeyearpar{tully77} relation linking galaxy luminosity to the circular velocity $V_c$ measured at large radii ($>R_e$), which is typically used as a proxy for dark matter halo mass. All of these dynamical scaling relations are remarkably tight and can be used to measure galaxy masses and distances (see \citealt{courteau14} for a review).

\begin{figure}[h]
\centering 
\includegraphics[]{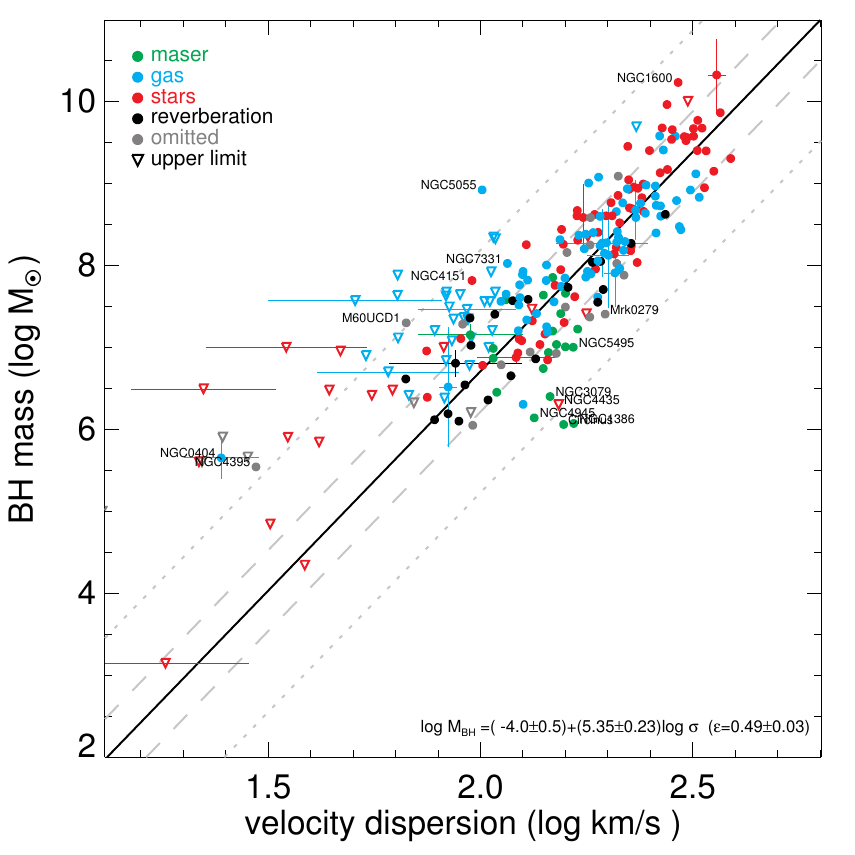} %
\caption{The tight correlation between black hole mass and stellar velocity dispersion. This solid line shows the \Msigma\ relation derived in \S\ref{sec:msigma}. It is based on \ngal\ galaxies spanning from dwarfs to brightest ellipticals. The scatter in this relation is \eqmsigmaeps . Upper limits are shown as open triangles. Different colors  denote different types of \Mbh\ measurement (\S\ref{sec:datamethods}). Error bars are only shown for the objects with the largest uncertainties. The grey dashed and dotted lines denote 1 and 3 times the intrinsic scatter.}
\label{fig:bhsigma} \end{figure}

Host galaxy properties also correlate with the mass of the super-massive black holes (BHs) in their centers (see \cite{kormendy13a} for a review). %
Out of all BH scaling relations (none of which is as tight as the Fundamental Plane), the so-called \Msigma\ relation has been firmly established as the strongest, most universal relation \citep{gebhardt03,beifiori12}.  This %
empirical relation between BH mass (\Mbh ) and velocity dispersion, shown in figure~\ref{fig:bhsigma}, has the least scatter of all other BH scaling relations.  %
It applies to galaxies of all types, including the most massive galaxies -- like M87 -- and dwarf galaxies, such as local group members M33  and NGC205. There are only few outliers to the \Msigma\ relation. In the compilation of galaxies studied in this paper, the most notable outliers are the $10^6$ \Msun\ BHs from water masers \citep[\S\ref{sec:itworks}]{greene10a,greene16, kuo11}.  %
The galaxy with the smallest BH is M33 (=NGC598). The putative BH in this object has an upper limit  on its mass of 1500 \Msun\ \citep{gebhardt01, merritt01}.  Considering the amount of scatter about the \Msigma\ relation, even these outliers are not especially far off the relation.

There is only one empirical relation tighter than the \Msigma\ relation, but this applies only to a small subset of galaxies: in the most massive bulge-dominated ellipticals, BH mass is correlated with bulge mass \citep[e.g.][]{haring04, gultekin09, kormendy11,lasker14a}. %
Since not all galaxies have a bulge, the \Msigma\ relation is more generally applicable.  The relevance of the empirical `black hole-bulge' mass relation is also questionable in the case of the \textit{pseudo}-bulges in disk galaxies,   %
which do not correlate with black hole mass in the same way \citep{hu09, jiang11a, graham13},  if at all \citep{sani11, kormendy11,lasker15,saglia16}.  It should be noted, though, that comparative bulge studies are complicated by the difficulties of bulge and disk photometric decompositions, which are notoriously degenerate \citep[e.g.][]{lasker14} and depend on non-unique definitions of what a bulge is \citep{kormendy04}.  %

\begin{figure}
\centering 
\includegraphics[width=8.7cm]{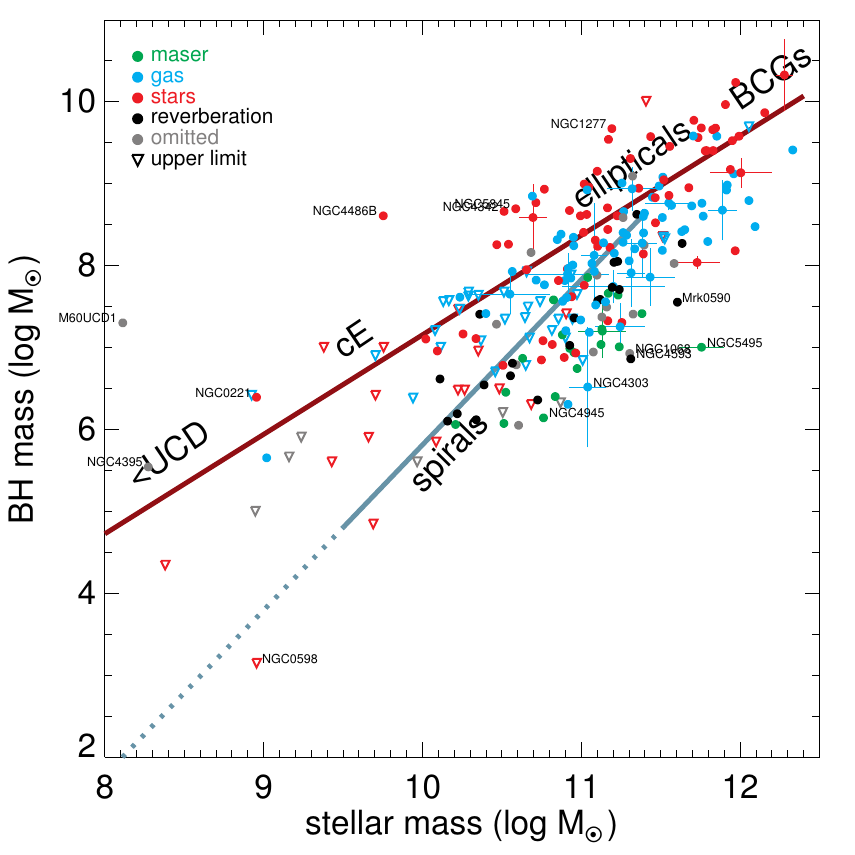}  
\caption{Total stellar mass and BH mass do not correlate well, as is shown here. The scatter of a simple regression (tab.~\ref{tab:msigma}) is \eqMbhLkeps, which is significantly larger than on the \Msigma\ relation. However the distribution of points make the interpretation as a single powerlaw difficult \citep{reines15}. Near $10^{11}$ \Msun\ the BH mass varies by 3 orders of magnitude.  Many different interpretations exist for subsets of galaxies and/or their bulges.  The (compact) ellipticals -- and classical bulges --  appear to follow the red line. And the low mass (disk) galaxies appear to follow the blue line. The two populations converge near $10^{11}$ \Msun . See section~\ref{sec:reprojections} for the different projections of these scaling relations. The symbols are the same as in figure~\ref{fig:bhsigma}.}
\label{fig:bhmass} \end{figure}

A much more compelling correlation would avoid the bulge altogether and link BH mass directly with total galaxy mass  (Figure~\ref{fig:bhmass}).  More massive galaxies do indeed have bigger black holes.  However, the manner in which observed objects populate $M_{\star}$-\Mbh\ space makes it difficult to claim that a single power law can accurately describe the correlation.  At host galaxy masses near $10^{11}$ \Msun\ the BH mass varies by 3 orders of magnitude. The AGNs appear to be offset from the ellipticals by an order of magnitude \citep{reines15}.  The over-massive black holes in bulge-dominated galaxies are also outliers \citep[e.g.  M60UCD-1, NGC1271, NGC1277 and NGC4486B;][]{seth14,walsh15,walsh16,saglia16}.  The only galaxies that appear tight are the ellipticals \citep{kormendy13a, lasker14a}.  Thus, all available evidence suggests that there is no clear  universal relationship between BH and galaxy mass.  The \Msigma\ arguably provides the much better predictor of BH mass. %

Several other empirical relations have been considered in the literature, but most of these are just manifestations of the simple rule that more massive galaxies have bigger black holes \citep[e.g. NSC mass, core radius, globular clusters, S\'ersic $n$, pitch angle:][]{ferrarese06,kormendy09,harris11,savorgnan13,berrier13}. None of these relationships are tighter -- or have less scatter -- than \Msigma\ over the whole mass range (see the review by \cite{graham15b} and references therein). Notably, the circular velocity does not correlate well with BH mass \citep{ho07, kormendy11a, sun13}, which is curious as it implies that the dark matter halo mass does not correlate with BH mass. 

So far there have been very few satisfying multi-variate studies of BH scaling relations.  %
Studies that consider additional parameters have mostly focused on the Black Hole Fundamental Plane\footnote{Perhaps the Black Hole Bulge Plane would be a better name?} by adding bulge parameters, like bulge size or mass \cite[e.g.][]{feoli09,marconi03a, hu09, aller07, hopkins07d, sani11, graham08, saglia16} to the  \Msigma\ relation. The most exhaustive multi-variate search was done by \cite{beifiori12}, who confirmed that the  \Msigma\ was the best single-parameter relation. They could only marginally improve it by adding $R_e$ as the secondary parameter. %

If \Msigma\ is the best single-variate relation, what does $\sigma$ truly represent? The total stellar mass does not correlate tightly with BH mass. Not all host galaxies have bulges.  Alternative interpretations for \Msigma\ must therefore be investigated. Here, I undertake a new multi-variate study of BH scaling relations and examine the link between \Msigma\ and the global photometric properties of BH host galaxies. Beginning with a section on the data \& methods, I describe the sample selection and the measurements of total luminosities and half-light radii. The regression fitting technique is described later in section~\ref{sec:regression}. Then, in section~\ref{sec:FP} I first show that, for the adopted sample, the BH host galaxies themselves lie on a tight Fundamental Plane.  I then confirm that the \Msigma\ is indeed the best single-parameter regression in section~\ref{sec:msigma}.  Section~\ref{sec:itworks} explores whether the \Msigma\ is internally consistent with the host galaxy Fundamental Plane identified in section~\ref{sec:FP} and then establishes the \BHLRname , using stellar masses estimated using the  mass-to-light conversion described in section~\ref{sec:stellarmass}.  Finally, I discuss the implications for BH scaling relations for different types of galaxies in section~\ref{sec:reprojections} and conclude in section~\ref{sec:conclusions}. Throughout, I adopt a flat concordance cosmology with $H_0=70$ \kms\ and $\Omega_m=0.3$. %

\section{Data and methods}
\label{sec:datamethods}

This section provides details on the construction of the sample of BH masses and host galaxy properties used in this work, as well as the methods employed for regression-fitting.  A brief summary is included below and in subsection~\ref{sec:photometrysummary}, which should be sufficient for most readers (who may want to skip the more technical details that follow).

To achieve proper leverage on the scaling relations, a large dynamic range in BH masses and host galaxy properties is essential. In this work this is achieved with a sample of \ngal\  galaxies compiled from the literature. For each object there is a black hole mass measurement in which the BH is either dynamically or temporally resolved. The galaxies and their properties are listed in table~\ref{tab:sample} and~\ref{tab:incomplete}. Additional parameters and code for generating the figures and fits are available at \url{https://github.com/remcovandenbosch/Black-Hole-Mass-compilation}. The BH masses in this compilation come from four different methods: stellar dynamics, gas dynamics, megamasers and reverberation mapping. Each of these methods uses a different dynamical tracer and probes a different region in the potential well of the galaxy \citep[e.g. ][]{peterson14a,kormendy13a,walsh13,vdb16}.  To highlight any potential biases that result, the symbol colour in all figures indicates the method used.  

The sample under consideration is a factor two larger than most previous compilations \citep[except][]{beifiori12}. This makes it possible to investigate regions of parameters space (\Mbh$<10^6$ \Msun\ and diffuse galaxies) that otherwise would not be probed.  The local group dwarf galaxies M33 and NGC205 especially help characterize the low mass end. 

The reader is cautioned that the mass measurements and their underlying assumptions -- especially pertaining to stellar dynamics --  are different for each measurement and, as a result, the sample is very heterogeneous. When different methods are applied to the same galaxy the results do differ \citep[for a discussion see][]{walsh13,vdb16}. In addition, the biases of detecting a BH mass as well as null-results are a potential issue \citep{ferrarese05, batcheldor10, gultekin11}.  Black hole masses can only be measured dynamically in galaxies that are extremely nearby and it is much easier to find bigger black holes than smaller ones \citep{vdb15}. Next generation facilities will significantly expand the parameter space \citep{davis14a, do14}.  For now, although sample heterogeneity will certainly impact the scatter on the resulting black hole scaling relationships, this is not of particular relevance for this work, which focuses on the exploring the connection between the different relationships.  As long as all fits are performed on the same sample they can be compared in a relative sense. 

\subsection{The Black Holes}
\label{sec:sample}

The basis of the sample consists of the 97 galaxies from the \href{http://blackhole.berkeley.edu}{online compilation (link)} by \citet{mcconnell13}. These are all objects hosting black holes whose masses are measured with some kind of dynamical models in which the kinematic tracer have been spatially resolved. Added to that are four objects from the \cite{gultekin09a} compilation, another four from the \cite{kormendy13a} compilation and fifteen from \cite{saglia16}. An additional 39 objects  have (updated) black hole mass (upper limits) from \cite{gebhardt01}, \cite{merritt01}, \cite{valluri05}, \cite{barth09}, \cite{kormendy10}, \cite{seth10}, \cite{neumayer12},
\cite{lyubenova13}, \cite{de-lorenzi13}, \cite{scharwachter13}, \cite{onken14}, \cite{nguyen14}, \cite{gultekin14}, \cite{yildirim15},
\cite{onishi15}, \cite{walsh15}, \cite{seth14}, \cite{walsh16}, \cite{barth16b}, \cite{thomas16}, \cite{greene16} and \cite{onishi16}. Another 80 objects with \textit{HST}/STIS spectroscopy are added from \cite{beifiori09,beifiori12}. The galaxies  from \cite{beifiori09,beifiori12} with low dispersions were shown to appear systematically offset from the \Msigma\ and hence I treat those objects with dispersions below 105 \kms\  as upper limits in this compilation.\footnote{I speculate that the $\sigma_e$ in the lowest mass galaxies of \cite{beifiori12} are underestimated, as they do not include the rotational part of the second moment \citep{falcon-barroso16}.   While their \Mbh\ measurements are consistent with the \BHLRname, their low dispersions make it appear that the BH masses lie above the \Msigma.  The Fundamental Plane would indeed predict higher dispersions.  See also \S\ref{sec:sample} and \S\ref{sec:FP}. \label{foot:B09}}   The remainder is included as BH mass measurements.

\textit{Temporally resolved black holes.} 
Also included in the sample are the 50 BH masses from \citet{bentz15}, for which the size of the broad emission-line region around the black hole is resolved using  time lags by repeated observations. Only 24 of these objects are included in the analysis. The remainder of the host galaxies are not sufficiently bright or large enough for a robust growth curve with the 2MASS data (\S~\ref{sec:photometry}). The BH masses are scaled using the virial factor $\langle f \rangle$ of 4.31$\pm1.05$ from \cite{grier13a}.\footnote{Note that the $\langle f \rangle$-factor is derived under the assumption that the AGN galaxies follow the \Msigma\ relation. As an independent consistency check of $\langle f \rangle$, I excluded all the reverberation mapped galaxies from the \BHLRname\ fit (from \S\ref{sec:itworks}) and then used the resulting regression to predict the black hole masses of the 26 reverberation mapped galaxies with growth curves. The weighted mean of the difference between that and the measured BH masses gives $\langle  \log f\rangle=0.62\pm0.04$. This is consistent with the adopted literature value.} 
The reverberation BH masses expand the sample towards lower black hole masses and shows that the AGN galaxies also follow the \BHLRname.

\textit{Velocity dispersions.}\label{sec:dispersions} For compatibility with the Virial Theorem, the ideal measurement constraining the dynamical state of galaxies would be the second moment of the 3D velocity tensor inside the half light radius. However this is not directly observable in external galaxies. The closest is the measurement of the stellar velocity dispersion of all the combined spectra  inside an aperture the size of the effective radius, as this is the 1D projection of the second moment. Inside the half-light radius the baryons dominate (See section \ref{sec:FP} and \ref{sec:stellarmass}). 

The measurement of $\sigma_e$ is most easily performed with an integral-field spectrograph \citep[IFU, e.g. SAURON,][]{emsellem04}, by collapsing the entire spectrum inside 1 $R_e$. However, IFU observations are not available for most of the sample.  Different studies have used different definitions of $\sigma$ to correlate with black hole mass. The widely used definition of $\sigma_e$ from \cite{gultekin09a} adopts the luminosity-weighted average of $\langle V^2+\sigma^2 \rangle$ per spatial kinematics bin inside one effective radius.  This includes both the rotational $V$ and the dispersion $\sigma$ component of the velocity tensor \citep[c.f.][]{bennert11,woo13}. This is often computed from long-slit observations \citep[e.g.][]{bellovary14}. The \Msigma\ relation study by  \cite{mcconnell13} uses a similar definition, but excludes the central region inside the BHs gravitational sphere-of-influence, where the black hole  influences the dispersion. Given that the BH is part of the self-gravitating system that the FP probes, this region should arguably be included in the computation of $\sigma_e$.

In the present compilation, for each galaxy the dispersion in the literature that most closely approximates $\sigma_e$ is used (where available). The literature dispersion are superseeded by the $\sigma_e$  measurements from \cite{cappellari13b}, \cite{saglia16}, \cite{mcconnell13} (in decreasing order of priority). When a dispersion is not otherwise available it is taken from \cite{grier13a},  \cite{vdb15}. Notably, good $\sigma_e$'s are not available for the lowest-mass spirals. These particular galaxies are so nearby that their  quoted literature dispersions are measured in the central arcseconds  only.  Since the half-light radius for these systems is more than an order of magnitude larger, these dispersions do not include the contribution of the disk rotation into the second moment. To convert the $\sigma$ of the lowest-mass spirals from \cite{barth09}, \cite{kormendy10}, and \cite{neumayer12} to $\sigma_e$, I adopt the aperture correction $\nicefrac{\sigma_e}{\sigma}=(\nicefrac{R}{R_e})^{0.08}$ for spirals from CALIFA IFU observations following \cite{falcon-barroso16}. For these spirals, the correction is about 20\% and is still relatively small with respect to the large dynamic range of the BH sample presented here. This correction is not applied to any other galaxies (see also footnote~\ref{foot:B09}).

The literature dispersions are all measured with different instruments and methods and  are often selected for study based on consistency with the relation \citep[c.f.][]{jardel11}. Surprisingly, though, the inhomogeneity of these dispersions does not seem to impact the results; if the dispersions are in error, they would not yield the very tight fundamental plane shown in section~\ref{sec:FP}.  Note that small changes in the aperture size do not affect the dispersion significantly: $\nicefrac{\sigma_e}{\sigma}=(\nicefrac{R}{R_e})^{\alpha}$, where $-0.06<\alpha<0.08$  \citep{falcon-barroso16}.  

\textit{Distances.} Where possible in the sample I adopt the galaxy distances and their errors from the compilation by \cite{saglia16}. The average distance uncertainty throughout that whole sample is 9\%. For all other galaxies distances are taken from their literature sources.  For these, an uncertainty of 10\% is adopted unless an uncertainty is specifically quoted.  The distances to local galaxies are relatively uncertain, due to the relatively large contribution of their peculiar velocities to the estimation of their redshifts, yielding high distance uncertainties. Unfortunately, redshift-independent distances are not available for most of these objects.

\subsection{Photometry}
\label{sec:photometrysummary}
In order to properly consider black hole masses in the context of the Virial Theorem -- c.q. Fundamental Plane -- global photometric properties are needed. The total near-infrared luminosity is a suitable proxy for the total stellar mass while the major axis length of the isophote containing half the light is a good proxy for half-light radius, as it is not strongly affected by projection effects.  Almost all  galaxies with a \Mbh\  are close and bright enough such that the $K_s$-band imaging from the Two Micron All Sky Survey \citep[2MASS,][]{skrutskie06} is sufficient.\footnote{Throughout the remainder of the paper $K_s$ is shortened to $K$.}  The $K_s$-band is not very sensitive to dust or to changes in stellar mass-to-light ratio \citep{bell01}.  In the infrared bands galaxy sizes are also smaller, i.e. than in optical bands, as a result of inside-out growth; at shorter wavelengths, the contribution from young optically bright stars leads to larger measured sizes.  

Using the growth-curve method detailed in the next section, I measured the photometry of the 260 galaxies from the full sample that are both resolved ($R_e>1$\arcsec ) and adequately detected ($M_k < 11$) in 2MASS. This is the same selection cut as the parent sample used in the HETMGS \citep{vdb15}.  They are listed in table~\ref{tab:sample}. Only 30 objects in the present compilation do not fit these criteria. 
 Many of these are  reverberation-mapped AGN, which are either too dim or are effectively point sources in the 2MASS photometry. The objects without photometry are listed in table~\ref{tab:incomplete}. In the other cases where the growth-curve analysis could not be performed, the catalog is augmented with photometric data from the 2MASS XSC \citep{jarrett00} or Large Galaxy Atlas \citep{jarrett03} or from the BH mass literature, when available.  These cases are recognizable in the data tables by the lack of uncertainties on their photometric values. 

The growth-curve analysis provides non-parametric determinations of the galaxy luminosities, sizes, and concentration indices (and their covariances). This is achieved by fitting each galaxy multiple times with 3 S\'ersic functions in which the outermost S\'ersic $n$ is varied. The S\'ersic function is selected for its convenience and ability to fit galaxy photometry extremely well.  %

\subsubsection{S\'ersic growth curves}

\label{sec:photometry}
\vspace{5mm}
\newcommand{\attrib}[1]{%
\nopagebreak{\raggedleft\footnotesize #1\par}}
\relsize{-1}
\settowidth{\versewidth}{Three S\'ersics to bring them all and in the darkness bind them}
\begin{verse}[\versewidth]
Three S\'ersics for the Elven-kings under the sky, \\
Seven for the Dwarf-lords in their halls of stone, \\
Nine for Mortal Men doomed to die,\\
One for the Dark Lord on his dark throne,\\
In the Land of Galaxies where the Shadows lie,\\
One S\'ersic for strong residuals, One S\'ersic to fiat them,\\
Three S\'ersics to bring them all and in the darkness bind them\\
In the Land of S\'ersic-fits where the Shadows lie.\\
\end{verse}
\relsize{+1}

\attrib{The Lord of the S\'ersics, epigraph}
\vspace{5mm}

In this section I describe the technical details of the photometric measurements. The 2MASS atlas images have dimensions of 512$\times$1024 pixels of 1\arcsec\ each with a  typical PSF size of 3\arcsec\ Full-Width-Half-Maximum (FWHM). To prepare for the S\'ersic fitting, all the images are pre-processed with  \texttt{SEXTRACTOR} \citep{bertin96}, to find and mask all the stars and extended objects. Then a global PSF is constructed using \texttt{PSFEX} \citep{bertin11} that extends 30 times beyond the FWHM, with a plate scale of 0.5\arcsec. The galaxies in the sample are much larger than the PSF, so fluctuations in the PSF core size are less relevant than the spiders and the coma of the PSF, especially when fitting the AGN cores and bright foreground stars. The central core of the PSF is governed by the subsampling of the 2\arcsec\ detector pixels onto the 1\arcsec\ output grid.  

\begin{figure}[b]
\centering 
\includegraphics[width=7.0cm]{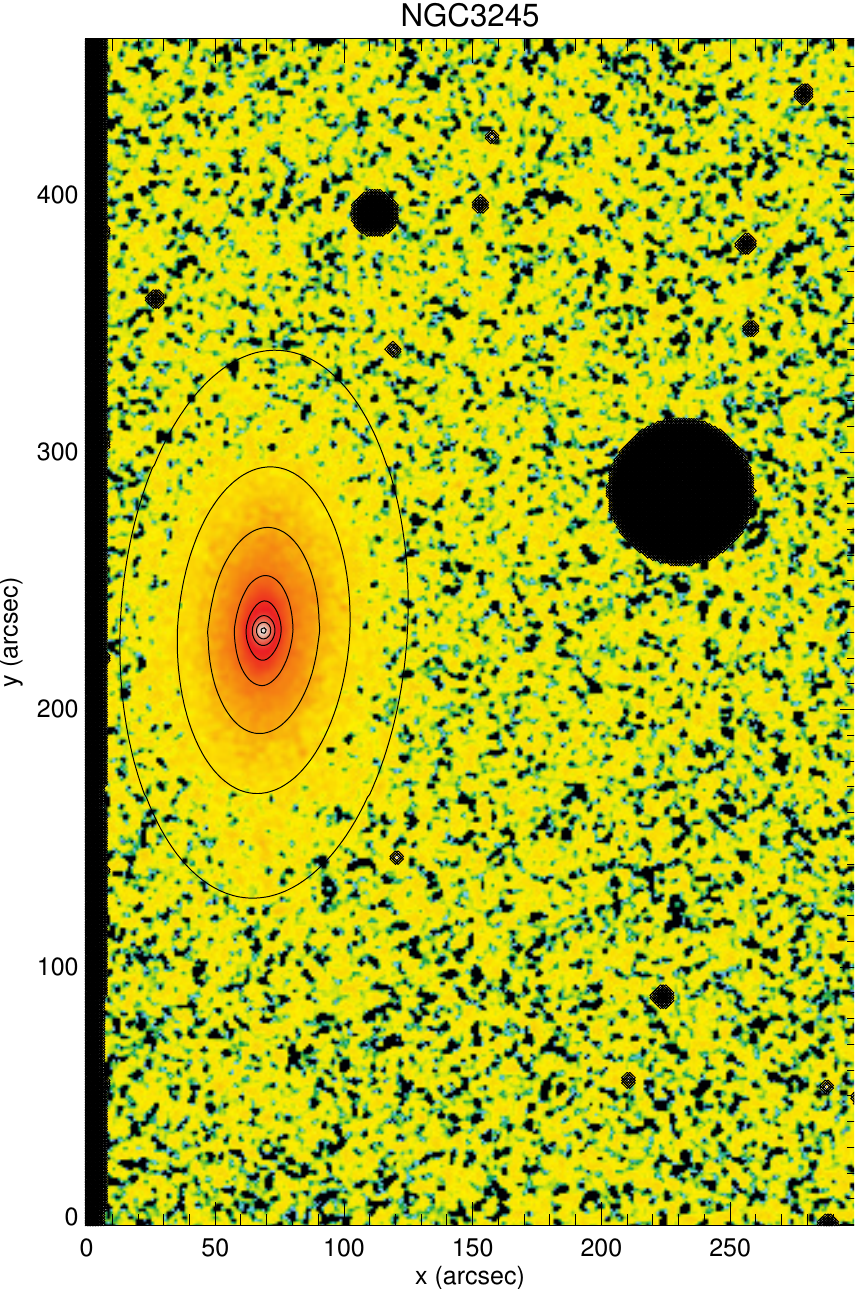} %
\caption{The 2MASS atlas image of NGC3245. Overplotted are the contours  of equal surface brightness for the best-fit model in steps of 1 magnitude. Distant bright stars and galaxies are masked. Nearby stars and galaxies are added to fit.  The images of the other galaxies  are available in a figure set.}
\label{fig:image} \end{figure}

Next, an initial galaxy model is made with \texttt{galfit} \citep{peng02} using the parameters from the 2MASS extended source catalog  \citep[XSC,][]{jarrett00} as seed values. Each fit has been manually inspected to ensure a reasonable convergence. The model contains three S\'ersics profiles for the galaxy, plus a central (non-thermal) point source (when an AGN is present) and a sky plane with variable tilt. Nearby bright stars and extended objects are unmasked and included while fitting the galaxy of interest as a point source and a single S\'ersic, respectively. The fit is performed on a region ten times larger than the extent of the target galaxy.  An example fit is shown in figure~\ref{fig:image}. 

Several additional constraints aid numerical convergence: the luminosities of each S\'ersic are required to be within 4 magnitudes of each other; all three S\'ersic functions are constrained to have the same centroid position and position angle; the S\'ersic indexes are reqired to be in the range $0.5<n<4$.  The choice of thee S\'ersic profiles is driven by the desire to have enough freedom in the inner part and at the same time have an outer S\'ersic for the growth curve. This appears to be the minimum required to fit the photometry without large residuals \citep{huang13}. %

\begin{figure}[b]
\centering 
\includegraphics[ width=0.85\columnwidth]{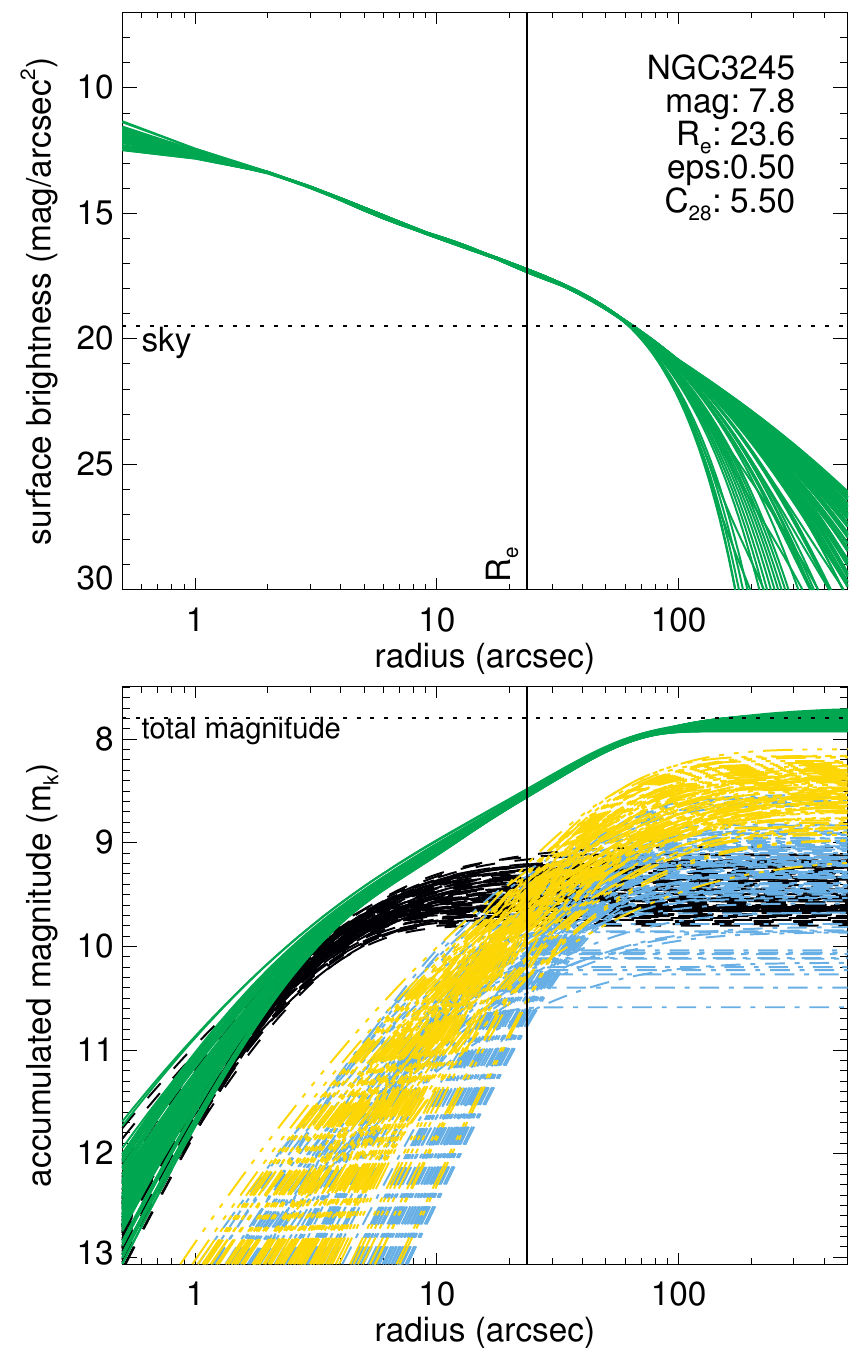}%
\caption{Example of the growth curve fit. The top panel shows the deconvolved surface brightness along the major axis for NGC3245. For each of the 100 fits, the outer S\'ersic index $n$ is changed from 0.5 to 4. Above the sky level (dotted line) the photometry is well constrained. Below the sky level,  varying $n$ forces the photometry to probe the largest extent of profiles with different slopes allowed by the data. The central region also shows  scatter, due to degeneracy with the AGN point source included in the model. The half-light radius is shown as the vertical line.  The bottom panel shows the curve-of-growth, radius versus enclosed light.  Even though the surface brightness profile is not very well constrained at large radii, the enclosed mass is very well constrained, as the outer part below the sky level does not contain a significant amount of flux. The three S\'ersic profiles of each iteration in Monte-Carlo fits are shown as the (dotted-)dashed lines. Multiple S\'ersic fits are very degenerate, however the total flux is well constrained. The extrapolated total magnitude is shown as the dotted line. Identical figures for the other galaxies are available in a figure set.}
\label{fig:S\'ersic} \end{figure}

Starting with the initial galaxy model, \texttt{galfit} is run 100 times for the subsequent growth curve analysis. In each fit, the starting conditions -- including the sky pedestal -- are varied by 10\% and the outer slope is varied by fixing the index $n$ of the outer S\'ersic from 0.5 to 4, in the 100 incremental steps. Additionally,  the half-light radius of the outer S\'ersic is forced to be at least 20\% larger than the other S\'ersics.  This ensures that the outer-slope of the model is fixed to the intended value.  This scheme essentially represents a `brute force approach' to fitting growth curves with different slopes at large radii.  As an example, the 100 surface brightness profiles of NGC3245 resulting from such a Monte-Carlo run is shown in figure~\ref{fig:S\'ersic}.

The combined iterations provide a measure of the uncertainties and covariance of the parameters of the surface brigthness profile.  The total flux in the 3 S\'ersics profiles represents the total galaxy luminosity. The half-light radius $R_e$ (=$R_{50}$) is the major axis length of the isophote that contains 50\% of the light in the deconvolved image. (Similarly, the radii $R_{20}$ and $R_{80}$ each contain 20\% and 80\% of the light, respectively.) To include uncertainty on the radius measurements, 0.5\arcsec\ scatter in the PSF FWHM is added\footnote{In practice this factor is negligible apart from the galaxies with the smallest apparent sizes, like NGC4486B.}.  Then the correlation matrix of the total luminosity, half-light radius and concentration $C_{28}=5\log R_{20}/R_{80}$ is computed. 

The uncertainties on the photometric properties are strongly correlated and are not negligible. The median uncertainty on the total luminosity and half-light radius are 0.09 mag and 0.06 dex. The magnitudes are converted to solar luminosity $L_K$ by adopting an absolute solar luminosity for the Sun of 3.28 magnitudes and a correction for the foreground extinction from \cite{schlegel98}. All the measurements used in the fits presented in this paper are in table~\ref{tab:sample}. Other numbers and covariances are \href{https://github.com/remcovandenbosch/Black-Hole-Mass-compilation}{available online}.

\begin{figure}
\centering 
\includegraphics[width=1.0\columnwidth]{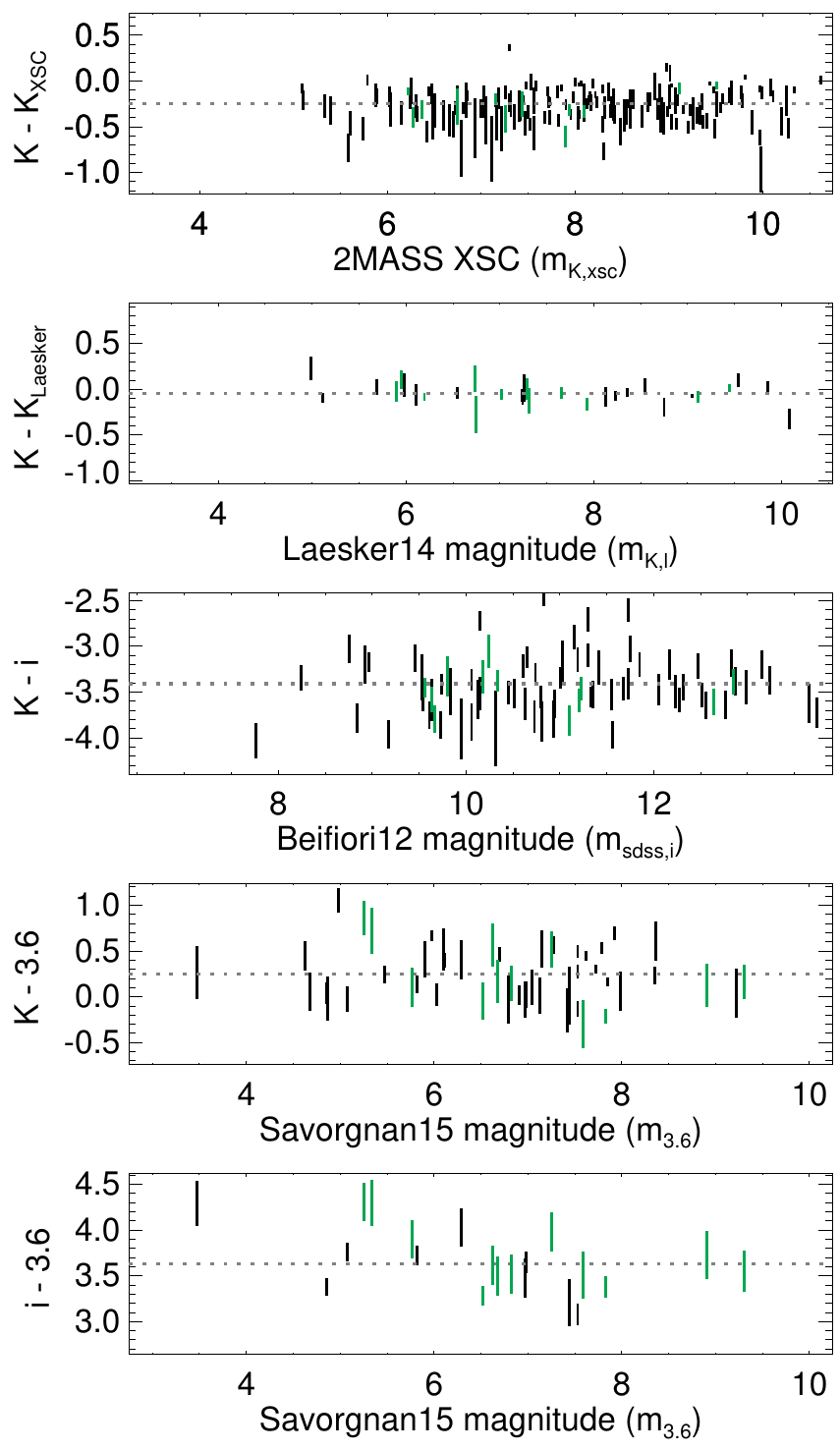} 
\caption{Comparison of the total apparent magnitudes with respect to other studies and bands. From top to bottom the comparison band becomes redder. The top shows the comparison between SDSS $i$ band from \cite{beifiori12} derived using isophotal analysis. The middle panel shows the total magnitude  derived from deep $K$-band using \texttt{galfit} from \cite{lasker14a}. The bottom panel shows the comparison with \textit{Spitzer} 3.6 micron from \cite{savorgnan16a}. The range on the $x$- and $y$-axis is the same in all panels.  Residuals are smallest when comparing identical $K$ band, as is expected. Mean colors are $m_K-m_{K,l\ddot{a}sker}= 0.05$, $m_K-m_i= -3.4$ and $m_K-m_{3.6}= 0.25$. While there is significant scatter (due to the different bands and methods), no systematic trend is seen as function of apparent magnitude. This indicates that the 2MASS images is deep enough to measure the total flux for these objects.  The green objects are common to all samples.}
\label{fig:photcomparison} \end{figure}

The Two Micron All Sky Survey is relatively shallow and thus the photometry does not resolve low surface brightness features.  The sizes ($R_{K\_R\_EFF}$) and apparent magnitudes ($m_{K\_M\_EXT}$) tabulated in the XSC, in particular, are known to be underestimated \citep[e.g.][]{lauer07,cappellari11}. Figure~\ref{fig:photcomparison} shows a comparison of total magnitudes measured here with values in the literature. Reassuringly, the photometry presented in this work is comparable to the very deep K-band photometry observed with CFHT/WIRCAM of 35 galaxies from \cite{lasker14}. As expected, there appears to be a significant offset between the growth-curves and the XSC. Using the  technique from section~\ref{sec:regression}, I find that the XSC magnitudes are significantly fainter than the 2D growth curves (i.e. the magnitude measured here $m_K =  -0.33+ 1.01 m_{K\_M\_EXT}$), with inferred errors of $\delta m_{K\_M\_EXT} =  0.18$. The XSC apparent sizes are also significantly smaller, as expected. The best-fit transformation from those sizes to the values measured here with the growth-curve analysis is $\log R_e =   1.16\log R_{K\_R\_EFF} +  0.23 \log q_{K\_BA}$, in which $q_{K\_BA}$ is the flattening of the galaxy in the XSC. Comparison with deeper photometry yields better results. The regression with the deep K-band photometry from \cite{lasker14}  is $m_K = 0.03  + 1.003 m_{K,l\ddot{a}sker}$ \cite[where $m_{K,l\ddot{a}sker}$ is the integrated magnitude inside the 24 mag/arcsec$^2$ isophote, as defined by][]{lasker14}, which indicates a perfect match within the uncertainties. The work by \citealt{lasker14} does not quote individual errors, nonetheless the inferred uncertainty of $\delta m_{24}  =  0.05$ is consistent with their estimated systematic error.  This indicates that the 2D growth curves on the 2MASS photometry is sufficiently accurate for the purpose of determining the black hole scaling relations and that photometry is not the limiting factor \citep[see also][]{vika12}. 

For reference, figure~\ref{fig:photcomparison} also shows a comparison of the apparent magnitudes measured in K band with those measured in SDSS $i$-band from  \cite{beifiori12} and with \textit{Spitzer} 3.6 micron from \cite{savorgnan16a}.  It should be noted that comparisons between different bands suffer from  differences in stellar populations and dust attenuation; even dust-free ellipticals have color gradients. In addition, total magnitudes from the literature are not derived using the growth-curve analysis adopted here. The $i$-band total magnitudes were derived using isophotal analysis and the 3.6$\mu$ magnitudes were measured in the context of Bulge-Disk decompositions. While I find a significant scatter of $\sim 0.3$ dex in the colors, there is no systematic dependance on total luminosity, indicating that the 2MASS images are deep enough to measure the total flux for these objects.

\subsection{Fitting linear regressions}\label{sec:regression}

Linear regressions can be fitted using various algorithms. The four most commonly used to measure BH  scaling relations are those that have been tested by e.g. \cite{park12a}. In this work, I use the \texttt{mlinmixerr} and \texttt{linmixerr} routines by \cite{kelly07}. These routines use a Bayesian approach to multiple linear regression. The main reason for choosing this Gibbs sampler is that it includes the ability to incorporate upper-limits on the black hole masses and co-variance between the observables.  Here the algorithm is used to find the coefficients of regressions of the form $Z=\alpha + \beta X + \gamma Y$  as well as the scatter $\epsilon$ around variable $Z$. In this work, the scatter is around the BH mass, except when applied to the galaxy Fundamental Plane in the next section, where it is on the $\sigma_e$ variable. The mean and the standard deviation of the posterior distribution supplies the quoted measures of the best-fit values and $1 \sigma$ uncertainties. 

The regressions in this work are all log-linear regressions. Thus all the BH masses, luminosities, sizes, dispersions and their errors are converted to logarithmic units before fitting.  The unit of $Re$ is kiloparsec,  while $L_k$ is given in solar luminosities, $\sigma_e$ is in \kms , and mass is in \Msun .   

The fitting algorithm allows for uncertainties on all parameters as well as for covariance between them. Uncertainties on the photometry and the spatially resolved BH masses are both affected by the distance uncertainty.  Although the distance is not an explicit parameter of the regression, the error needs to be well-known and correctly propagated to the uncertainties of the other variables by combining them in quadrature. For most galaxies, the uncertainty on the BH mass is significantly bigger than the distance error. The velocity dispersion is not affected by distance errors, as the aperture in which it is measured (should) depend only on the apparent size of the galaxy. Luminosity, half-light radius and \Mbh\ all depend on distances and thus cause (additional) covariance between parameters.    

The most significant covariance is between the luminosity (c.q. mass) and size, which is directly computed during the Monte-Carlo growth-curve fits. This includes the effect of the distance error onto the covariance. For the scaling relations fits in this paper, the other covariance terms are very small and are (assumed to be) negligible \citep{saglia16}.

\subsection{Omitted Objects}\label{sec:omitted}

\label{sec:rejects} All regressions in this paper are performed on the same set of \ngal\ galaxies to ensure that they can be reliably compared against one another. Not all galaxies in the compilation are included in the fits.  As previously mentioned, 32 galaxies are excluded because of incomplete data (Table~\ref{tab:incomplete}). (They lack either 2MASS growth curves (\S\ref{sec:photometry}) or  a velocity dispersion.)  Another 15 galaxies are excluded from the fits because they are strong systematic outliers and one of more of their data are arguably suspect.  Specifically, under the assumption that the scaling relations presented here are true, these 16 galaxies are outliers because: 

\begin{itemize}
  \item The literature dispersion is too high (NGC1300) or low (NGC2139, NGC5194, NGC4041) or has a unrealistically small error. %
   \item Both NGC2139, NGC4636, NGC5018, NGC4699, NGC4826 and NGC4736 are significant outliers in the FP and \BHLRname , possibly due to systematic uncertainties in their distances.  The latter three BH masses come from unpublished literature  \citep[see][]{kormendy11,saglia16}.  %
\item Both NGC1068 and NGC7469 are much too bright in $K$-band. They appear denser and brighter than the densest galaxies in the sample. Either their distance is in error \citep{yoshii14} or there is significant additional flux from the starburst seen in these galaxies \citep{wilson86}. 
  \item MRK110, MRK279 and NGC3156 have a bright AGN and its host galaxy is too faint for a robust growth curve with 2MASS. 
  \item  The surface brightness of NGC4395 is too low for a robust growth curve analysis with 2MASS. Based on optical photometry, the half-light radius is 1 kpc and the total stellar mass is $10^{8.9}$ \Msun\ \citep{den-brok15,reines15}, which is consistent with their black hole mass and velocity dispersion estimates.
\end{itemize}

All of the above cases are systematically excluded from all fits considered in this study. However, for full disclosure, they are all still included in the figures but marked as `omitted'. Including (excluding) individual objects  increases (decreases) the measured scatter, but does not  significantly change the coefficients of the regressions. Previous studies often exclude more objects, which is why they typically find smaller scatter on their scaling relations. For example, \url{http://blackhole.berkeley.edu} lists 97 galaxies, of which 19 are labeled `complicated'.

\section{The galaxy Fundamental Plane} \label{sec:FP}

The first step in the analysis is the construction of a Fundamental Plane for all the black hole host galaxies. The ETGs are well known to obey a tight Fundamental  Plane. Spiral galaxies also obey dynamical scaling relations.  Although they are not typically included in the Fundamental Plane, they too must obey the Virial Theorem as the underlying gravitational physics is the same. Applied to ETGs, the FP works so well because, inside the half light radius, the gravitating mass consists almost completely of stars; dark matter, the BH and non-homology are not dominant factors  inside 1 $R_e$ (see~\ref{sec:itworksMass}) .  In addition, the collision-less nature of stars makes them a good tracer of the second moment of the velocity tensor. The same is also true for star-forming disk galaxies. The dark matter fraction inside the half-light radius of a spiral galaxy is small, with the dark matter only becoming dominant at two disk scale lengths \citep{bershady11, courteau15}, which is equivalent to 3.3 $R_e$ for an exponential disk. The main complication for spirals is that the light more often (than in their early-type counterparts) traces a mixture of stellar populations of various ages and metallicities as well as star formation. This can be mitigated by using photometry in the infrared where changes in mass-to-light ratio are minimized \citep[see also \S\ref{sec:stellarmass},][]{falcon-barroso11a,meidt14,norris14}.  The fit of a Fundamental Plane to large samples of heterogeneous galaxy types has already been successfully considered \citep{zaritsky08, falcon-barroso11a,bezanson15}.

As described below, the black hole host galaxies in the present sample, with uniform $K$-band photometry, do indeed follow the Fundamental Plane. In an effort to stay reminiscent of the Virial Theorem, I choose to consider the form $\sigma_e \propto L_k^\alpha R_e^\beta$.  This has the benefit that 1) it has nearly the same variables as the canonical black hole scaling relations 2) the variables on both sides are (nearly) independent from one another and 3) the \texttt{mlinmixerr} routine only supports upper limits on the fitted variable, which will become relevant later. 

The regression results in 

\begin{multline}
	\label{eq:FP}
\eqfp .
\end{multline} 

\begin{figure}
\centering 
\includegraphics[]{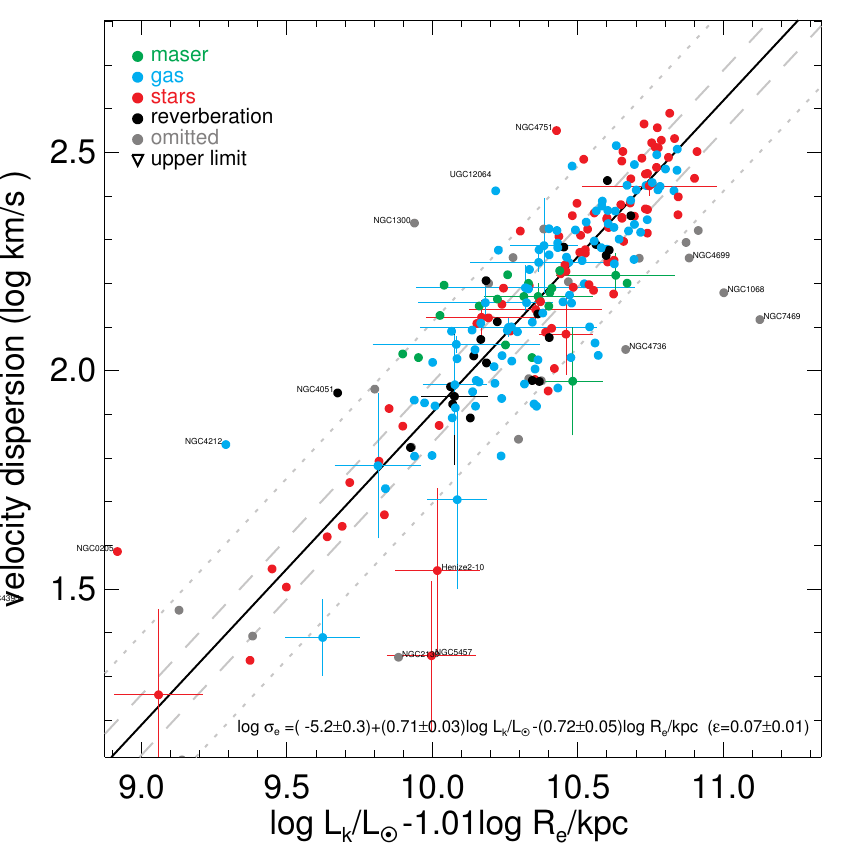} %
\caption{Galaxies lie on a remarkably tight Fundamental Plane with a scatter of \eqfpeps\  -- 5 times smaller than the \Msigma relation. This demonstrates that the galaxy luminosity and size together are an extremely good proxy for the stellar velocity dispersion (see~\S\ref{sec:FP}). Note that objects outlying the relation most likely have issues with their data and are not suspected to be true outliers of the Virial Theorem.  Symbol shape is the same as in figure~\ref{fig:bhsigma}, while symbol color denotes the type of \Mbh\ measurement (\S\ref{sec:datamethods}) performed for each host galaxy (although BH mass is not a fitted parameter).  Note as well that the $x$-axis has the same units as in figure~\ref{fig:bhmasssize}. }
\label{fig:sigmamasssize} \end{figure}

\noindent  The edge-on view of this plane is shown in figure~\ref{fig:sigmamasssize}. The intrinsic scatter is only 
\eqfpeps . (Note that there are a few outliers interpreted as bad data, see \S\ref{sec:omitted}) and excluded from this and subsequent fits.) The scatter is very small, especially when compared to the black hole scaling relations for which the scatter is  six times larger.  Adding a concentration term, such as S\'ersic $n$ or $C_{28}$, improves the mass estimator by a small amount \citep[e.g.][]{zaritsky06,courteau07a,taylor10b,courteau14}, but such second order effects are outside the scope of the current paper. 

The coefficients of the FP regression in eq. \ref{eq:FP} are remarkably close to the Virial Theorem, yielding \fpLk . This corresponds to a `tilt' with respect to the Virial Theorem by \fptilt . Within the uncertainties, this tilt is solely in the $\sigma_e$ direction. The half-light radius term is negligible, as also found by \cite{cappellari13b} for a sample of 260 early-type galaxies.  The regression also provides a constraint on the covariance between the two exponents on $L_k$ and $R_e$, implied to be quite strong strong and negative (-0.9).  This suggests that the largest amount of freedom in the tilt is predominantly in the $\sigma_e$ direction.  Overall, the small tilt and the tightness of the FP provide strong indications that this Fundamental Plane is very close to the Virial Theorem.  The $K$-band luminosity thus appears to be a very good proxy for the baryonic mass for these galaxies. Later in section~\ref{sec:stellarmass} the tilt in the Fundamental Plane will be discussed and used to convert the total luminosity to total stellar mass.  %

From figure~\ref{fig:sigmamasssize} it can be concluded that there is an empirical Fundamental Plane in the $K$-band that applies to all the black hole host galaxies with very little scatter. This plane holds equally well from the lowest mass $10^8$  \Msun\ dwarf galaxies to the $10^{13}$ \Msun\ BCGs. The existence of this tight Fundamental Plane provides validation of the dataset, including the homogeneous photometry, the stellar velocity dispersion and the distances. In addition, since many of the galaxies probed here do not have a bulge, this suggest that $\sigma_e$ is not intrinsically linked with the bulge mass but rather more directly to global photometric properties, as the strong correlation in figure~\ref{fig:sigmamasssize} implies.   Although bulge fraction might still be found to link with the Fundamental Plane parameters, the prominence of the bulge seems most likely to show a (partial) dependence on the concentration of the galaxy. %

\section{The M$_\bullet$--\,$\sigma$ and higher order relations} \label{sec:msigma}

Now that the Fundamental Plane of the host galaxies is established, this section focuses on the black hole scaling relations. The first is the relation between black hole mass and velocity dispersion. With the sample of black hole host galaxies defined in section~\ref{sec:sample}, the resulting regression for the \Msigma\ relation is 

\begin{equation}
\eqmsigma ,
\label{eq:msigma}
\end{equation} 

\noindent which is shown in figure~\ref{fig:bhsigma}. This is quite similar to that found in previous studies, which is to be expected given that they consider the same galaxies. The regression is specifically consistent with \cite{mcconnell13} and \cite{saglia16} but slightly steeper than \cite{beifiori12, kormendy13a}. The intrinsic scatter around \Mbh\ here, which is \eqmsigmaeps\, is $\sim0.1$ larger than found in previous work. This is primarily due to the increased sample size, which is twice as large as most previously studied samples and includes many more low mass objects.  Note, though, that the scatter could of course be artificially reduced by selectively excluding the largest outliers, as is typical of previous results. The scatter may also be sensitive to (small) differences in the adopted errors and in the fitting method, which varies between this and other analyses. The bayesian \texttt{linmixerr}  fitting routine used here is more general and produces a larger scatter with respect to other methods \citep{park12a}. %

A tightening of the \Msigma\ relation might be possible with the introduction of a secondary photometric parameter, as considered by \cite{saglia16}.  That recent, exhaustive study explored several higher order regressions in detail and found only a minor improvement over the single variate \Msigma\ relation when including bulge parameters. Other searches integrating the photometry of the bulges have identified relations with much stronger dependence on $R_e$ and a low exponent on $\sigma_e$ such that \Mbh $\appropto \sigma_e^{3} R_{bul}^{0.4}$ \cite[e.g.][]{marconi03a, aller07,hopkins07d,saglia16}. But none of these relations are (significantly) better than the \Msigma\ relation.  Especially when considering the large systematic uncertainties inherent to the heterogeneous sample.

As tabulated in table~\ref{tab:msigma}, here I consider whether global host properties rather than bulge properties can serve as an additional tightening parameter and fit regressions between $\sigma_e$ and either $R_e$, $L_k$ or concentration $C_{28}$.  None of these additional parameters yields a significant improvement, or even changes the slope on $\sigma_e$ outside of the $1\sigma$ uncertainty.  I thus conclude that the \Msigma\ is a universal relation spanning from the smallest to largest BH masses, is independent of galaxy type,  and has no dependence on any additional global photometric parameter. This reaffirms the earlier conclusions of \cite{gultekin09} and \cite{beifiori12} that the black hole scaling relation is driven solely by the velocity dispersion.

\begin{table}[h]
\centering
\relsize{-1}
\begin{tabular}{cclclc}
\hline
$\alpha$ & $\beta$ & X & $\gamma$ & Y & $\epsilon$ \\
\hline
\eqmsigmatab
\eqmsigmaRetab
\eqmsigmaLktab
\eqmsigmaCtab
\eqMbhMtab
\eqMbhLktab
\hline
\eqbhlrtab
\eqbhmrtab
\eqmfixlrtab 
\hline
\end{tabular}
\relsize{+1}
\caption{Black hole scaling relations based on parameter X and optionally a second parameter Y.  The form of the regressions are $\log$ \Mbh$  = \alpha +\beta $X$ + \gamma $Y with intrinsic scatter $\epsilon$. Adding a second parameter does not improve the scatter, nor significantly change the coefficients on \Msigma\ relation.
\label{tab:msigma}} 
\end{table}

\section{Unification of the Fundamental Plane and Super-Massive Black Holes masses}

If the \Msigma\ relation (eq.~\eqref{eq:msigma}) obeyed by black holes does not depend on any additional parameters and the galaxies that host them follow the Fundamental Plane (eq.~\eqref{eq:FP}; with only a small tilt in the $\sigma_e$ direction), then it follows that there is a black hole scaling relation in which $\sigma_e$ can be replaced by $(\nicefrac{L_k }{R_e})$. This is indeed the case, as I show in this section.  

\subsection{Constraining $\sigma$ and \Mbh ~with galaxy luminosity \& size} \label{sec:itworks}
\begin{figure*}[t]
\centering 
\includegraphics[width=11cm]{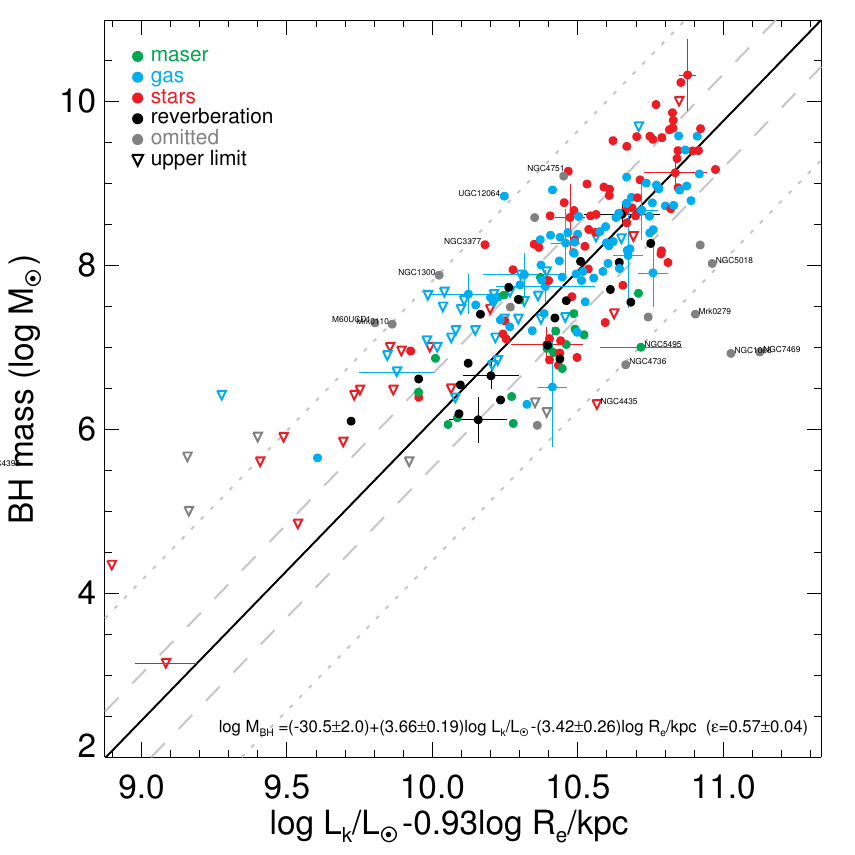} 
\caption{The \BHLRname\  -- shown here -- has a similar amount of scatter as the \Msigma\ relation. The black-hole--plane tilts in the same direction as the galaxy Fundamental Plane (fig.~\ref{fig:sigmamasssize}). Hence the relation shown here is just a reprojection of the \Msigma\ relation. The coefficients of this regression have nearly the same ratio as the Fundamental Plane, which is why the $x$-axis here is very similar to the one in figure~\ref{fig:sigmamasssize}.  The symbols are the same as in figure~\ref{fig:bhsigma}.}
\label{fig:bhmasssize} \end{figure*}

To demonstrate that, as expected from the Fundamental Plane, $\sigma_e$ can be replaced by $(\nicefrac{L_k }{R_e})$ in the \Msigma~relation, I first show that there is a strong correlation between \Mbh ~and $(\nicefrac{L_k }{R_e})$ by fitting an independent regression.  Then I show that this relation is consistent with the \Msigma\ relation.  Figure~\ref{fig:bhmasssize} shows this independent fit for the \BHLRname\ of the form 

\begin{multline}
\eqbhlr ,
\label{eq:bhlr}
\end{multline}

\noindent with an intrinsic scatter of \eqbhlreps . Remarkably, the coefficients on the $\log{L_k}$ and $\log R_e$ terms are of the same magnitude even though they are independent variables of the fit. Moreover, as these coefficients are of opposite sign, when considered as exponents on $L_k$ and $R_e$ they imply that the ratio $L_k/R_e$ is fundamentally well constrained, i.e. equation~(\ref{eq:bhlr}) implies \bhmasssize . This would seem to lend credibility to the $(\nicefrac{L_k}{R_e})$ identity, as it did for the FP in section~\ref{sec:FP}.  Imposing a 1D regression with $\log (\nicefrac{L_k}{R_e})$ as the sole variable yields a nearly identical coefficient, as tabulated in  table~\ref{tab:msigma}, and with almost the same amount of scatter and  uncertainty as equation~(\ref{eq:bhlr}).  This form \bhmasssize\ has the advantage that it has one less free parameter and the covariance between $L_k$ and $R_e$ becomes implicit. 

Most importantly, the regression given in equation~(\ref{eq:bhlr}) is equatable to the empirical \Msigma\ relation.  With the help of the galaxy Fundamental Plane, eq.~\eqref{eq:FP},  it can be recast as 
\begin{equation} \label{eq:MbhLR}
\fancyeq ,
\end{equation}
\noindent  which is consistent with the independently-derived \msigmaapproxerr\  from section~\ref{sec:msigma}. This shows that luminosity--size regression is, in fact, consistent with being a projection of the \Msigma\ relation. This consistency is also evident from the nearly identical $x$-axes shared by the galaxy Fundamental Plane in figure~\ref{fig:sigmamasssize} and the \bhmasssize\ regression in figure~\ref{fig:bhmasssize}. I conclude that this \BHLRname\ is just another identity of \Msigma.  

This was similarly found by \cite{beifiori12} -- namely \Mbh$\propto L_{i,\star}^{3.15\pm0.57} R_e^{-2.76\pm0.64}$ -- albeit with larger uncertainties.  In other studies, such as \cite{barway07} and \cite{hopkins07d}, the focus was on (decomposed) photometry of bulges and hence similar coefficients are not found.  As bulges lie on a narrow sequence  in mass and size (the Kormendy relation, see \S\ref{sec:reprojections}), it can be difficult to robustly fit them with a plane.  

Interestingly, whereas the Maser galaxies appear as slight outliers in the \Msigma\ relation \citep[fig.~\ref{fig:bhsigma}),][]{greene16} this is less pronounced in the \BHLRname. Many of the objects that have BH mass upper limits in the literature are actually consistent with the \BHLRname . It appears that the literature dispersions for many of these objects may be a little off; the Fundamental Plane also predicts a different dispersions ($\sigma_e$) for these objects (see also footnote~\ref{foot:B09}).

\subsection{Constraining $\sigma$ and \Mbh ~with galaxy mass \& size} \label{sec:itworksMass}
So far in this work I have considered only empirical scaling relations based on directly observable quantities, such as the total luminosity of the galaxy. However, the total stellar mass is a much more convenient basis for comparison with other works and theoretical predictions. The $K$-band luminosity is already a very good proxy for stellar mass, because variations in stellar mass-to-light ratios are strongly reduced in the \mbox{(near-)infrared} bands compared to optical bands  \citep{meidt14,norris14}.  Color information can generally be used to reliably estimate the mass-to-light ratio, but it is worth noting that dynamical and spectroscopic models both imply that the massive ellipticals have poorly understood IMF variations that are not manifest in broad band colors \citep{smith14a}. 

\subsubsection{Conversion to stellar mass} \label{sec:stellarmass}

For the sake of simplicity, I use yet another scaling relation to estimate the mass-to-light ration $M_\star/L_k$. Recent studies of the Fundamental Plane indicate that roughly 50-100 percent of its tilt is caused by variations of the stellar mass-to-light ratio \citep{cappellari06,cappellari13a,zaritsky06,bolton08,auger10,graves10,falcon-barroso11a} and the remainder due to non-homology, projection effects, the black hole and dark matter.\footnote{The super-massive black hole itself could be a contributor to the tilt of the FP.  The BH mass fraction is typically much less than 1\% and does typically not contribute significantly to the global observables of a galaxy. The main exception is M60UCD1 in which the BH mass fraction is bigger than 10\%. This stripped object is not included in the fits, because it is not resolved in the 2MASS imaging \citep{seth14}. In second place is NGC4486B (7\%) which is regularly considered to be an outlier \citep{gultekin09a}. In third place is NGC1277 (2\%). The latter two objects are included in the FP and the \Msigma\ relation in this work and are not an outliers. However, simple numerical tests do show that the BH can contribute to the dispersion in extreme cases. Isotropic Jeans models \citep{watkins13} of NGC1277 and M87 (without dark matter) with and without a BH show a change in $\sigma_e$ of about nine percent. The BH could thus be a partial contributor to the tilt in the fundamental plane.} Hence I adopt the average and assume that 75\% of the tilt is solely caused by mass-to-light variations and that galaxies with a dispersion of 166~\kms\ have  $1.0 $ \MLsunK\ from \cite{kormendy13a}. This directly yields \MLfunc. This factor, which is required only to convert $L_k$ into $M_\star$, represents only a small change (less than a factor 2) in the mass-to-light ratio across the full range in $\sigma_e$ probed by the galaxies in the sample.  %

Using this conversion, the Fundamental Plane can be converted into a virial mass estimator: $\VTestimator$. The slight (assumed) tilt is only in the velocity dispersion direction. This estimator and the ad hoc factor $\kappa=\VTkappa$ have been studied in detail in many other works (see \citealt{courteau14} for a review). The value of $\kappa$ found here is slightly higher than the value found by \cite{cappellari16} due to the (assumed) offset of $\sigma_e^{-0.14}$ with respect to the Virial Theorem.

\subsubsection{An independent regression in terms of galaxy mass}

The \BHLRname\ can be also re-fit with galaxy mass replacing K-band luminosity to yield 
\begin{multline}
\eqbhmr ,
\label{eq:bhmr}
\end{multline}
\noindent (The scatter is slightly lower than the original due to inclusion of the $\sigma_e$ in the definition mass-to-light ratio used to convert $L_k$ to galaxy mass. See table~\ref{tab:msigma}.) Just like the \BHLRname, this mass relation can be expressed in terms of only two constants instead of three, as \bhMRapprox, given the curious fact that the tilt of the Fundamental Plane is in the same direction as the \Msigma\ relation.  This form is especially useful when comparing to the predictions of semi-analytical models and to, e.g., the sub-grid physics in hydro-dynamical simulations of galaxies and the accretion histories of the super-massive black holes. %

It should be emphasized that the main conclusion of this paper  --- \msigmaapprox $\equiv$ \bhmasssize\  --- is an empirical relation and is completely independent of the choice for the mass-to-light ratio conversion.

\section{Black holes in relation to the sizes and masses of disks and bulges} \label{sec:reprojections}

\begin{figure}[t]
\centering 
\includegraphics[]{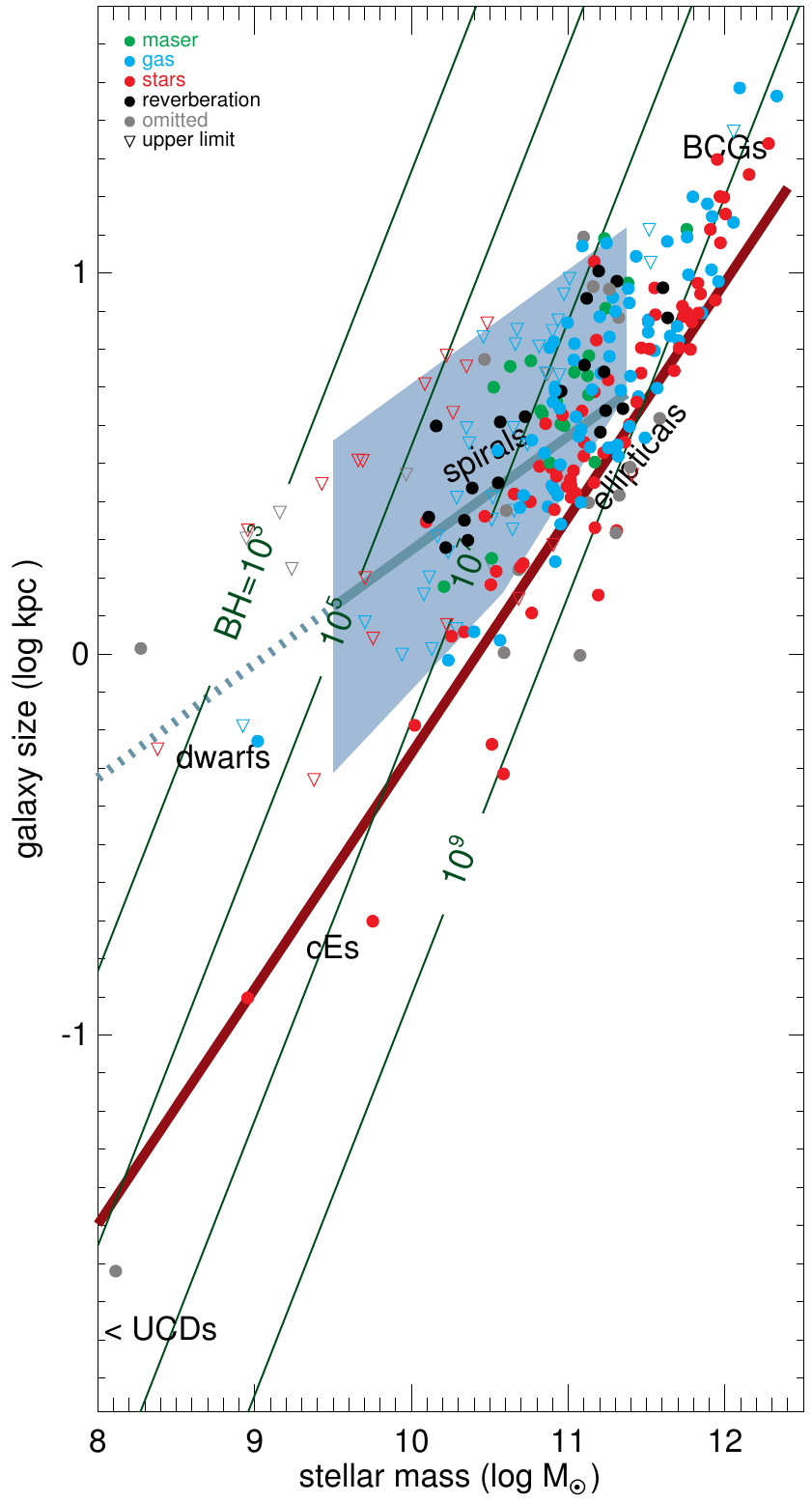} %
\caption{Distribution of the sizes and total luminosities of the sample with respect to the spiral and elliptical galaxies. The ellipticals fall on the Kormendy relation, shown as the red line. The late-type galaxies also follow a much wider mass--size relation, shown in blue. At $10^{11}$ \Msun\ the two families intersect. A 1D projection of these two relations is shown in figure~\ref{fig:bhmass}. The \BHMRname\ is plotted as green contours, indicating the expected black hole mass. This plots highlights the uneven sampling of the black hole host galaxy distribution which mostly reflects the limits imposed by the spatial resolutions of current telescopes. Color coding of the symbols is the same as in figure~\ref{fig:bhsigma}.}
\label{fig:sizemass} \end{figure}

Different subsets of galaxies occupy different regions of the Fundamental Plane. The distribution of the present sample of black hole host galaxies in mass and size is shown in figure~\ref{fig:sizemass}.  (Another view is shown in figure~\ref{fig:bhmass}.)  Overplotted in figure~\ref{fig:sizemass} are the mass--size relations of the ETGs and the disks galaxies. For other variations of this plot see \cite{janz16} for the distribution of quiescent galaxies or see \cite{van-der-wel14} for the size evolution of galaxies as function of redshift. The ETGs lie on a narrow sequence \citep{chen10c} and the disk galaxies lie on a relation with a larger scatter \citep{courteau07}. Where the two populations intersect is typically where lenticular galaxies are located.  In the next two subsections, the projections of these two populations will be discussed separately.  

As revealed in this study, the black hole scaling relations are all linked together though the Fundamental Plane.  The predicted BH mass expected from the \bhmasssize\ relation is indicated relative to the galaxy population by contours in figure~\ref{fig:sizemass}. Compared to previous compilations \citep[e.g.][]{shankar16}, the significantly expanded sample of \ngal\ objects considered here is much more representative of the galaxy population.\footnote{The \cite{shankar16} sample was published during the review of this paper.} Still, the host galaxies in the present sample are not very homogeneous across the global galaxy population.  Some regions are over-sampled while others are very sparse.  The $L_\star$ dense ellipticals, for one, are over-represented \citep{yu02a, vdb15, shankar16}, whereas the low mass galaxies are under-sampled.  More leverage on the black hole scaling relation can be created by targeting host galaxies in the under-sampled regions.  The lowest mass galaxies are underrepresented because of the resolution limit of 0.1\arcsec\ of today's optical and near-infrared telescopes \citep{vdb15}

\subsection{Ellipticals and Bulges} \label{sec:Kormendybulgebh}

Black holes are well known to exhibit a strong correlation with bulge mass, but at first glance the \bhmasssize\ and the canonical \Mbh $\propto M_{bul}^1$ relation appear incompatible.  As described below, however, this is just a projection effect.  

Consider, to start, that ellipticals (and their bulges) obey a tight relation between their size and mass \cite[e.g.][]{kormendy09a}.  This reprojects into the Faber-Jackson \citeyearpar[$L\propto \sigma^4$,][]{faber76} and Kormendy \citeyearpar{kormendy77a} relations using the Fundamental Plane.  The same power-law followed by ellipticals is also followed by compact ellipticals and UCDs \citep[e.g][]{norris14a}, all systems with a Bulge-to-Total ratio that is close to unity (i.e. almost  all of their mass is in a bulge).  In this case, $M_\star\sim M_{bul}$ and, as such, the relation shown in red in Figure~\ref{fig:sizemass} for the densest galaxies, can be translated into a relation in terms of bulge mass.  Figure~\ref{fig:sizemass} specifically shows the
mass--size relation from  deep K-band photometry from  \cite{lasker14}.   Their 34 bulge-dominated galaxies yield an independently derived  relation: 

\begin{equation} \label{eq:LRetg}
\bulgeRL
\end{equation}

\noindent All of the galaxies close to this (red) line in figure~\ref{fig:sizemass} are bulge-dominated \citep[e.g.][]{savorgnan16a} and thus have Bulge-to-Total ratios of nearly 1 so that $M_\star\sim M_{bul}$.  Equation~\eqref{eq:LRetg} can thus be combined with ~\eqref{eq:MbhLR} to produce the black hole mass projection for these galaxies: 
\begin{equation}
\bulgeMM ,
\label{eq:etgbhbulge}
\end{equation} %

\noindent which is shown in figure~\ref{fig:bhmass}. This projection is consistent with other previous $K$-band \Mbulge\ relations \citep{kormendy13a,graham13,lasker14a}. It appears that the canonical $M_\bullet \propto M_{bul}^1$ relation is just the projection of the  \Msigma\ relation onto the Kormendy relation. Is this a coincidence? Or is it regulated by physics?  The principal question is what causes the slope of the quiescent galaxy mass--size relations, which is unchanging with redshift \citep{van-der-wel14}.

Although BH mass can be cast in terms of bulge mass, as in eq. \ref{eq:etgbhbulge}, the \Msigma\ relation, and its proxy  \bhmasssize , should be preferred over the $M_\bullet \propto M_{bul}^1$ as they are more universally applicable and do not require a bulge definition of any kind.  Both apply to all stellar systems and are completely independent of galaxy type and bulge-to-total fraction. Thus they also work for completely bulge-less galaxies. In contrast, the \Mbulge\ relation only applies to galaxies with a classical bulge. In the case of pseudo-bulges, the bulge mass is already well known to provide a less accurate prediction of black hole mass than in the \Msigma\ relation \citep{saglia16,savorgnan16}.  Furthermore, as shown in section~\ref{sec:msigma}, the \Msigma\ relation does not appear to be improved upon by, e.g. adding the $C_{28}$ concentration parameter.  

There is no existing 2D relation that includes bulge properties that has been found to be significantly better than the \Msigma\ relation.  It therefore seems plausible that there is, in fact, no causal link between the bulge and black hole mass.  Of course, revealing such a link, if one exists, is complicated by the difficulty of estimating the bulge fraction, which depends strongly on the definition of bulge used and is strongly degenerate when measured with multiple S\'ersics functions.  Even so, the scaling relations all appear to be tightly linked. For example, the $M_\bullet \propto (\sqrt{M_{bul}}\sigma_e^2)^{1.09}$ from \cite{saglia16} for classical bulges \citep[see also][]{hopkins07a,hopkins07b,aller07,feoli11} directly maps onto the canonical  $M_\bullet \propto M_{bul}^1$ relation when combined with the \Msigma\ relation.  For (pseudo-)bulges, which form a separate Fundamental Plane different from the total mass Fundamental Plane, a (crude) mapping between the total galaxy FP and the bulge FP also arguably exists, given that this plane still uses the same $\sigma_e$ \citep{sani11, saglia16}.  The multitude of secular processes thought to make (pseudo-)bulges grow in disk galaxies could somehow correlate with SMBH accretion, leading to a correlation between BH and pseudo-bulge mass. However, this process, if it exists,  does not seem to have the same vector as $\sigma$ on the Fundamental Plane.  

Despite the link between different scaling relations, the ~\Msigma\  and \bhmasssize\ relations do not necessarily always imply the same BH masses as the \Mbulge\ relation.  A prime example is the case of the black holes in the Brightest Cluster galaxies (BCGs), which are the most massive galaxies at the tip of the luminosity function.  When extrapolating from the lower mass ellipticals, these galaxies curve away from the Kormendy relation such that they have relatively large size for their mass \citep{lauer07b}.  As a result, their velocity dispersions hit a ceiling at 400 \kms.  The \Msigma\ relation thus predicts a lower BH mass than the \Mbulge\ relation.  (Note that massive galaxies are well represented in this compilation, with 31 galaxies more massive than M87, suggesting that the \Msigma\ relation as measured here remains valid at the high mass end.)  
Likewise, the \bhMRapprox\ predicts that the biggest black holes reside in the highest dispersion galaxies and not in the most luminous.  Lower mass ($10^{11.2}$ \Msun) galaxies with high dispersions are more numerous and likely host the biggest black holes \citep{lauer07}.

\subsection{Disk and spiral galaxies}

Disk galaxies also follow a mass--size relation, although it is not as tight as the Kormendy relation for bulge dominated ETGs.  For example, using the transformation from section~\ref{sec:photometry}, the size--luminosity relation from \cite{courteau07} based on the 2MASS XSC of spirals with $L_k>10^{9.5}$ \Msun\ can be converted to %

\begin{equation} \label{eq:LRltg}
\diskRL
\end{equation}

\noindent  This is shown as the blue line in figure~\ref{fig:sizemass}.  The shaded blue area denotes the $2\sigma$ scatter as measured by \cite{courteau07}.  (Note that the spiral galaxies in the present sample are reasonably sampled down to a mass of $10^{9.5}$ \Msun, but the sampling becomes very sparse at the low mass end.)  The mass--size relation for disks  can then be combined with equation~\eqref{eq:MbhLR} to predict a 1D black hole mass relation. The prediction is show as a blue line in figure~\ref{fig:bhmass} and can be expressed as

\begin{equation} \label{eq:diskMM}
\diskMM 
\end{equation} 

\noindent  This relation intersects with the ellipticals because the two mass--size relations also intersect. This is unlike the parallel relation for AGNs from \cite{reines15}, perhaps suggesting that the AGN host galaxies follow a different mass--size relation than the spirals.  

Low mass dwarf galaxies also host black holes. This is supported by evidence of active nuclei in their centers \citep[e.g.][]{greene07b,reines11,reines13,sartori15}. The dwarfs may very well be the best way to study the seed formation of SMBHs in the local universe. The smallest SMBH is found in RGG118, which has a black hole mass of $10^{4.7}$ \Msun\ \citep{baldassare15}.  The existence of a seed mass for SMBH formation would imply that the BH scaling relations need to truncate somewhere, in contrast to equation~\eqref{eq:diskMM}, which predicts that BH masses do not suddenly truncate.  One way to test whether such a truncation exists is to confirm that galaxies below a certain mass do not host  AGNs. For example, if the SMBH seed mass is $10^4$ \Msun, then eq.~\eqref{eq:diskMM} indicates that there would be no AGNs in disk galaxies with  masses below $10^{8.5}$ \Msun .

\subsection{Globular clusters and Ultra-Compact dwarfs}\label{sec:compactgal}

There has been much recent effort to measure the black hole masses in dense stellar systems, including Globular Clusters and Ultra-Compact Dwarfs (UCDs). These systems are not included in this compilation as they have very different formation histories than normal galaxies. Their black hole masses are also much more uncertain.  However, in this section, I briefly compare their BH mass estimates to the \BHLRname . 

Globular clusters are not quite dense enough to lie on the Kormendy relation \citep{misgeld11}. According to the \bhmasssize\ ~relation, given their sizes and masses these systems should not host intermediate mass BHs over 1000 \Msun . I speculate that Globular Clusters do not host SMBHs at all.  This is consistent with non-detections at radio and X-ray wavelengths \citep{strader12, miller-jones12,haggard13} and upper-limits from stellar dynamics \citep{vdb06,  lanzoni13,den-brok14, kamann14, kamann16}, but in contrast with the black hole mass detections from \citep{lutzgendorf11,lutzgendorf13,lutzgendorf15}. Yet other observational evidence \citep{lanzoni13,bianchini15} as well as arguments from dynamical analyses \citep[anisotropy and mass-to-light variations][]{vdb06,kamann16,zocchi15} suggest that this mass could be considered an upper limit.  The appendix~\ref{sec:GCtable} contains a compilation of BH mass measurements in GCs. Some of the Ultra Compact Dwarfs (UCDs) are dense enough to host SMBHs \citep{mieske13}. This is indeed the case for M60-UCD1 \citep{seth14}, but does not appear to be the case for NGC4546-UCD1 and UCD3 \citep{norris15,frank11}. Other UCD's are stripped galaxies that still host the original SMBH of their progenitor \citep{norris14a}.  The exact delineation between UCDs and globular clusters is not yet clear and many objects lie in between.  For instance, it is not clear whether M31's G1 \citep{gebhardt05} and $\omega$ Centauri \citep{vdv06,noyola10,van-der-marel10} are stripped galaxies, like the UCDs, or like the highest mass globular clusters (without a black hole).  Given their stellar masses and sizes they could be either.

\section{Discussion \& Conclusions} \label{sec:conclusions}

In this paper I examine the nature of the scaling relations linking black holes with their host galaxies.  For this empirical study I combine \ngal\  black hole masses and host galaxy velocity dispersions from the literature with new $K$-band photometry derived uniformly for all galaxies from 2D growth curves. With these data I confirm that BH host galaxies obey two empirical dynamical scaling relations: the Fundamental Plane \fpLk\ and the \msigmaapprox . The Fundamental Plane, traversing the space defined by galaxy mass, size and velocity dispersion, is the tighter of the two relations.  The 1D scaling relations obeyed individually by the dwarf, spiral and elliptical galaxy populations are thus recognizable as projections of the different distributions of each class of object on the FP.  

The fact that the Fundamental Plane applies to all galaxy types, including those with and without (pseudo-)bulges, has important implications for the mutual relevance of the relations that link BH mass to either the velocity dispersion $\sigma_e$ or to bulge mass.  Since the global photometry of bulge-less galaxies is a very good predictor of $\sigma_e$, the velocity dispersion by itself is arguably not a (direct) tracer of the bulge mass, as it is often invoked.  (Bulge mass correlates with the global host galaxy properties but can not be predicted solely by $\sigma_e$.)  Furthermore, only the \Msigma\ relation is relevant for the bulge-less galaxies that host super-massive black holes, like NGC4395 (and does indeed predict black hole masses consistent with observations). I therefore argue that the \Msigma\ relation is the optimum universal relation.  Even if the physics of bulge formation and black hole growth are linked, it thus appears that bulge properties are not an optimal predictor of black hole mass, particularly when a galaxy lacks a detectable bulge.  

The Fundamental Plane and the \Msigma\ relation together constitute a basis that can define other scaling relations applicable to galaxies of all types. In particular, I reveal the existence of the \BHLRname\ \bhmasssize , which has the same amount of scatter as the \Msigma\ relation. This scaling relation is completely expected from the combination of the \Msigma\ relation and the Fundamental Plane.  Thus, the \BHLRname\  is  just another identity of  \msigmaapprox, as I demonstrate in section~\ref{sec:Kormendybulgebh}, where I show that  the projection of the Kormendy relation for bulges is consistent with the canonical black-hole--bulge-mass relation of $M_\bullet \propto M_{bul}^1$.  

The  \BHLRname ~can be recast in terms of stellar mass, adopting a conversion between luminosity and stellar mass.  In section~\ref{sec:stellarmass} I adopt a simple mass-to-light ratio estimator based on the slight tilt observed in the Fundamental Plane with respect to the Virial Theorem.  The \BHLRname ~becomes \bhMRapprox\ and the total mass of a galaxy can be approximated with $\VTestimator$.

The new \BHLRname  ~(and its cousin, the \BHMRname) serves particularly well when the stellar velocity dispersion of the host galaxy is not known, such as in cases when the galaxy is faint (e.g. $z>0$ or low surface brightness), or when the  velocity dispersion is below the instrumental resolution.  The existence of the relation also offers a unique perspective on why total galaxy mass (or proxy $L_k$) by itself is not a good predictor of BH mass \citep[fig.~\ref{fig:bhmass}, ][]{reines15}, as has also been reproduced by recent cosmological galaxy simulations with SMBHs \citep[e.g.][]{angles-alcazar13,schaye15,steinborn15,volonteri16}.  If the \Msigma\ relation is indeed universal, the number density of SMBHs  \citep{lauer07,shankar09,kelly12a} would seem best derived from the galaxy velocity dispersion function \citep{sheth03,bezanson12}, rather than the galaxy mass function.  

At larger distances ($z>0$), where it becomes much harder to accurately measure $\sigma_e$, the \BHMRname\  is ideal.  It remains unclear, though, whether today's \Msigma\ relation also holds at earlier times.  Strong evolution is implied by the increase in galaxy size within the quiescent galaxy population with redshift \citep{van-der-wel14}, but there are hints that individual galaxies may still evolve such that the \Msigma\ relation stays intact at each epoch.   %
For example, the relic galaxies NGC1271, NGC1277, b19 and MRK1216 \citep{walsh15, walsh16,lasker13,yildirim15} all formed at $z>2$ and have evolved passively ever since \citep{ferre-mateu15}.  These galaxies are much smaller than most present day ETGs and host  extremely big black holes \citep{walsh15,walsh16}. Even though they formed a long time ago, they are still consistent with the black hole scaling relations today.  %

The scaling relations presented in this work can be improved in several ways. First, homogeneity in the measurements should be pursued. While the photometry all comes from the same instrument, the stellar velocity dispersions and black hole masses for the galaxies in the present sample are measured with a variety of techniques and instrumentation and are highly heterogeneous in quality.  Velocity dispersions measured from IFUs have been shown to significantly decrease the scatter in the FP \citep{cappellari13a},  
but such measurements are not available for most of these black hole host galaxies. 
The most homogenous spectroscopic dataset of BH hosts comes from the long slit observations from the HETMGS. Even with a homogenous dataset one must still insure that dispersions are defined and measured consistently. Different definitions of, and alternative units for, $\sigma$ will directly change the coefficients of the \Msigma\ relation and can change the scatter. In this work, I have preferred measurements of $\sigma_e$, as they are  the most widely available and the closest to the dispersion as expressed in the Virial Theorem. However, alternative definitions are also possible (\S\ref{sec:dispersions}) and might possibly yield a more fundamental \Msigma\ relation. For example, the definition  of the dispersion advanced by \cite{mcconnell13} specifically excludes the region around the BH, to decouple the measured dispersion from the influence of the BH.

 Another improvement can be made by using a stellar dynamics to forward model projection effects \citep{bellovary14} and derive dynamical mass-to-light ratios and the dark matter content with a homogenous data set and dynamical methods.  The non-uniformity of the black hole mass measurements themselves must also be alleviated.  The commonly used set of techniques for black hole mass measurements all probe different types of host galaxies, but ideally a significant sample is studied under the same conditions and with the same consistently applied technique.  This includes adopting the same set of assumptions in  stellar dynamical models, which have significantly changed over the last two decades \citep[e.g. triaxiality, dark matter haloes, and systematics, see][]{kormendy13a}. Cross-calibrations of BH mass measurement techniques and uniformity across larger samples of BH masses is highly desirable for future progress \citep{vdb16}. 

Although this paper describes several empirical black hole scaling relations, it does not examine their causality.  
The fact that the \Msigma\ relation appears to be universal and applies to all galaxy types would seem to hint at some form of feedback process between the SMBH and the host galaxy. Many of the most common theories for black hole--galaxy coevolution link black hole and bulge growth together.  Other theories involving direct black hole feedback link the black hole mass to $\sigma^4$ or $\sigma^5$, based on energy or momentum driven winds \citep{king03,fabian99,silk98}. The functional form found here, \msigmaapproxerr , is closest to the momentum driven theories.  Even if these theories are correct, the merging history must play an important role in establishing the \Mbh--host galaxy relation \citep{peng07, jahnke11} through the virtue of the central limit theorem %

\section*{Acknowledgments}

I am grateful for comments on the manuscript from the anonymous referee, the manuscript editor Sharon E.~M.~van der Wel, Stephane Courteau, Nathalie Ouellette, Lisa Steinborn, Glenn van de Ven, Akin Y{\i}ld{\i}r{\i}m, Tim de Zeeuw and, to Nadine Neumayer for insisting the figures should use stellar mass, instead of luminosity.

This publication makes use of data products from the Two Micron All Sky Survey, which is a joint project of the University of Massachusetts and the Infrared Processing and Analysis Center/California Institute of Technology, funded by the National Aeronautics and Space Administration and the National Science Foundation.

The work depended greatly on the public databases, \href{http://leda.univ-lyon1.fr}{HyperLeda} \citep{paturel97},  NASA's Astrophysics Data System and the NASA/IPAC Extragalactic Database (NED), which is operated by the Jet Propulsion Laboratory, California Institute of Technology, under contract with the National Aeronautics and Space Administration. This research has made use of NASA's Astrophysics Data System.

 \begin{center}
\begin{longtable*}{p{0.4in} ccc ccc cc c p{1.0in}}
 \hline
 \hline 
 Name & Distance &   \Mbh      & $\sigma_e$ &   $L_k$     & $R_e$  &   covar.         & $C_{28}$ & AGN        & method & ref.  \\
              &   Mpc        & log(\Msun )& log(\kms ) &  log(\Lsun) & log(kpc) &  $L_k$--$R_e$ &           & log(\Lsun)  &        &       \\ 
 (1)  &    (2)    & (3)        & (4)        & (5)        & (6)      & (7)            & (8)      & (9)        &  (10)&    (11)  \\ 
\\  \hline   \\                         
 \endfirsthead 
 \hline  
 Name & Distance &   \Mbh      & $\sigma_e$ &   $L_k$     & $R_e$  &   covar.         & $C_{28}$ & AGN        & method & ref.  \\
              &   Mpc        & log(\Msun )& log(\kms ) &  log(\Lsun) & log(kpc) &  $L_k$--$R_e$ &           & log(\Lsun)  &        &       \\ 
 (1)  &    (2)    & (3)        & (4)        & (5)        & (6)      & (7)            & (8)      & (9)        &  (10)&    (11)  \\
\\  \hline   \\                         
 \endhead
 \hline \\  \multicolumn{11}{r}{{Table continues on next page.}} %
 \endfoot  
 \hline \hline
 \endlastfoot
                       3C120 &  141.4$\pm$14.1 &               7.73$\pm$0.15 &   2.21$\pm$0.05 &  11.20$\pm$0.09 &   1.00$\pm$0.06 &            0.05 &     3.8$\pm$0.2 &    11.39$\pm$0.00 & reverb &                                                  {\tiny \cite{kollatschny14}} \\
                     3C390.3 &  240.3$\pm$24.0 &               8.62$\pm$0.16 &   2.44$\pm$0.03 &  11.25$\pm$0.10 &   0.64$\pm$0.09 &            0.14 &     3.9$\pm$0.8 &    11.26$\pm$0.04 & reverb &                                                     {\tiny \cite{dietrich12}} \\
                    A1836BCG &   152.4$\pm$8.4 &               9.57$\pm$0.06 &   2.46$\pm$0.02 &  11.75$\pm$0.06 &   0.89$\pm$0.06 &            0.04 &     4.9$\pm$0.5 &                   -  & gas &                                                  {\tiny \cite{dalla-bonta09}} \\
                    A3565BCG &    49.2$\pm$3.6 &               9.11$\pm$0.07 &   2.51$\pm$0.02 &  11.83$\pm$0.07 &   0.98$\pm$0.06 &            0.04 &     5.3$\pm$0.2 &                   -  & gas &                                                  {\tiny \cite{dalla-bonta09}} \\
                      Ark120 &  140.1$\pm$14.0 &               8.05$\pm$0.17 &   2.28$\pm$0.02 &  11.20$\pm$0.09 &   0.74$\pm$0.07 &            0.05 &     3.9$\pm$0.4 &    11.53$\pm$0.00 & reverb &                                                   {\tiny \cite{doroshenko08}} \\
                      Arp151 &    90.3$\pm$9.0 &               6.65$\pm$0.16 &   2.07$\pm$0.01 &  10.62$\pm$0.09 &   0.45$\pm$0.13 &            0.10 &    10.9$\pm$3.3 &                -  & reverb &                                                        {\tiny \cite{bentz10}} \\
                    Circinus &     2.8$\pm$0.5 &               6.06$\pm$0.10 &   2.20$\pm$0.05 &  10.22$\pm$0.12 &   0.18$\pm$0.13 &            0.09 &     6.7$\pm$0.6 &                 -  & maser &                                                    {\tiny \cite{greenhill03}} \\
\end{longtable*}
 
 \end{center}
 
 \begin{center}
 \begin{longtable*}{p{0.4in} ccc ccc cc c p{1.0in}}
 \hline
 \hline 
 Name & Distance &   \Mbh      & $\sigma_e$ &   $L_k$     & $R_e$  &   covar.         & $C_{28}$ & AGN        & method & ref.  \\
              &   Mpc        & log(\Msun )& log(\kms ) &  log(\Lsun) & log(kpc) &  $L_k$--$R_e$ &           & log(\Lsun)  &        &       \\ 
 (1)  &    (2)    & (3)        & (4)        & (5)        & (6)      & (7)            & (8)      & (9)        &  (10)&    (11)  \\
\\  \hline   \\                         
 \endfirsthead
 \hline 
 Name & Distance &   \Mbh      & $\sigma_e$ &   $L_k$     & $R_e$  &   covar.         & $C_{28}$ & AGN        & method & ref.  \\
              &   Mpc        & log(\Msun )& log(\kms ) &  log(\Lsun) & log(kpc) &  $L_k$--$R_e$ &           & log(\Lsun)  &        &       \\ 
 (1)  &    (2)    & (3)        & (4)        & (5)        & (6)      & (7)            & (8)      & (9)        &  (10)&    (11)  \\
\\  \hline                          
 \endhead
 \hline \\ \multicolumn{11}{r}{{Table continues on next page.}} %
 \endfoot  
 \hline \hline
 \endlastfoot
  \multicolumn{11}{c}{{Objects omitted due to incomplete data}} \\ \\ \hline \hline
      {\tiny 1RXSJ1858+4850} &           338.3 &               6.68$\pm$0.19 &              -  &              -  &              -  &              -  &              -  &                -  & reverb &                                                          {\tiny \cite{pei14}} \\
                   H0507+164 &           76.57 &               6.69$\pm$0.36 &              -  &              -  &              -  &              -  &              -  &                -  & reverb &                                                       {\tiny \cite{stalin11}} \\
                     M60UCD1 &           16.50 &               7.30$\pm$0.15 &   1.82$\pm$0.03 &            8.29 &           -1.62 &              -  &              -  &                  -  & star &                                                         {\tiny \cite{seth14}} \\
                          MW &        0.008000 &               6.63$\pm$0.05 &   2.00$\pm$0.09 &              -  &              -  &              -  &              -  &                  -  & star &                                             {\tiny \cite{ghez08,gillessen09}} \\
                     Mrk0006 &    80.6$\pm$8.1 &               8.09$\pm$0.15 &              -  &  11.11$\pm$0.09 &   0.66$\pm$0.08 &            0.06 &     6.0$\pm$0.6 &    10.99$\pm$0.01 & reverb &                                                   {\tiny \cite{doroshenko12}} \\
                     Mrk0142 &           192.5 &               6.27$\pm$0.21 &              -  &              -  &              -  &              -  &              -  &                -  & reverb &                                                           {\tiny \cite{du14}} \\
\end{longtable*}
 
 \end{center} 
 
\bibliography{bhfpcitekey}

\appendix
\section{Black hole measurements in globular clusters}
This appendix contains table~\ref{tab:GCsample} that lists a compilation of BH mass measurements in globular clusters.

\label{sec:GCtable}
\begin{table}[!h]
\centering
\relsize{-1}
\begin{tabular}{l ccc ccc ccc c}
\hline
 Name & Distance &   \Mbh      & $\sigma_e$ &   $L_k$     & $R_e$  &   covar.         & $C_{28}$ & AGN        & method & ref.  \\
              &   Mpc        & log(\Msun )& log(\kms ) &  log(\Lsun) & log(kpc) &  $L_k$--$R_e$ &           & log(\Lsun)  &        &       \\ 

\\  \hline   \\                         
                          G1 &          0.7300 &               5.24$\pm$0.12 &   1.39$\pm$0.09 &              -  &              -  &              -  &              -  &                    -  & GC &                                                     {\tiny \cite{gebhardt05}} \\
                     NGC0104 &        0.004000 &      3.00$^{+0.18}_{-3.00}$ &   1.06$\pm$0.03 &   5.80$\pm$0.19 &  -2.57$\pm$0.26 &            0.10 &     4.4$\pm$0.8 &                    -  & GC &                                                  {\tiny \cite{mclaughlin06a}} \\
                     NGC1851 &         0.01210 &      0.00$^{+3.30}_{-0.00}$ &   0.97$\pm$0.02 &   5.49$\pm$0.08 &  -2.77$\pm$0.06 &            0.05 &     4.5$\pm$0.3 &                    -  & GC &                                                  {\tiny \cite{lutzgendorf13}} \\
                     NGC1904 &         0.01290 &      3.48$^{+0.12}_{-3.48}$ &   0.90$\pm$0.03 &   5.25$\pm$0.10 &  -2.42$\pm$0.10 &            0.05 &     4.8$\pm$0.6 &                    -  & GC &                                                  {\tiny \cite{lutzgendorf13}} \\
\hline
\end{tabular}
\relsize{+1}
\caption{BH mass measurements in Globular Clusters. The columns are the same as table~\ref{tab:sample}.  Photometry for NGC 5272 is from 2MASS LGA \citep{jarrett03}, the remainder is derived using the growth-curves.
\label{tab:GCsample}} 
\end{table}

\end{document}